# Title: Lunar Swirls Unveil the Origin of the Moon Magnetic Field

**Authors:** **Boxin Zuo*[1], Xiangyun Hu[2], Lizhe Wang[1], Yi Cai[1] , Mason Andrew Kass[3]**

**Affiliations:**

[1] China University of Geosciences, School of Computer Sciences, Wuhan 430078, China.

[2] China University of Geosciences, Institute of Geophysics and Geomatics, Wuhan 430074, China.

[3]Aarhus University, Department of Geoscience, Denmark, Aarhus

Boxin Zuo*[1]: boxzuo@cug.edu.cn, ORCID 0000-0002-7274-1081

Tel: +86-13657263837

Address: No.388 Lumo Road, Wuhan, China

**Abstract:**

The origins of the lunar magnetic anomalies and swirls have long puzzled scientists. The prevailing theory posits that an ancient lunar dynamo core field magnetized extralunar meteoritic materials, leading to the current remnant magnetic anomalies that shield against solar wind ions, thereby contributing to the formation of lunar swirls. Our research reveals that these lunar swirls are the result of ancient electrical currents that traversed the Moon's surface, generating powerful magnetizing fields impacting both native lunar rocks and extralunar projectile materials. We have reconstructed 3-D distribution maps of these ancient subsurface currents and developed coupling models of magnetic and electric fields that take into account the subsurface density in the prominent lunar maria and basins. Our simulations suggest these ancient currents could have reached density up to 13 A/m², with surface magnetizing field as strong as 469 μT. We propose that these intense electrical current discharges in the crust originate from ancient interior dynamo activity.

**One Sentence Summary:** The origins of lunar magnetic fields and swirls have been quantitatively modeled and revealed.

# Main

Orbital magnetic data show that the lunar crust has been strongly magnetized by intense magnetic fields. Rock samples from the Apollo missions indicate that a global magnetic field (50–100 μT), generated by a core dynamo mechanism, existed in ancient times (4.2–2.5 billion years ago) (Garrick-Bethell et al. 2009; Le Bars et al. 2011; Shea et al. 2012; Weiss & Tikoo 2014; Buffett 2016, Mighani et al. 2020). However, the magnetism inherent in endogenous lunar materials is too weak to have recorded the core field. The prevailing theory suggests that extralunar impact meteorites with high thermoremanent susceptibility are the primary source of the strong remanent magnetization anomalies detected in the lunar crust (Wieczorek et al., 2012, 2023; Wakita et al. 2021; Citron et al., 2024).

Nevertheless, the origins of some lunar magnetic anomalies remain elusive. For instance, isolated strong anomalies in the maria, which are not associated with any known impact events, correlate closely with lunar swirl optical anomalies. It has been hypothesized that lunar swirls may form from the crust's magnetic field due to uneven space weathering from magnetic deflection of incoming solar wind (Hood & Schubert, 1980; Garrick-Bethell et al., 2011) or through hybrid and kinetic plasma interactions (Zimmerman et al., 2015). However, the fine and complex morphology of these swirls challenges our current understanding of their origins. Recent theories propose shallow, narrow iron-rich magmatic dikes or lava tubes as potential sources of the swirls (Hemingway & Tikoo, 2018). Yet, the magnetism of endogenous lunar materials is inadequate to produce anomalies as strong as those observed at orbital altitudes. Lunar Prospector, GRAIL, Chang'e and other orbiters have comprehensively measured the Moon's global magnetic vector field and gravity field, significantly enhancing the development of related models. These datasets are invaluable, containing critical information about the origins of lunar crust magnetization. In this study, we have reanalyzed these datasets from diverse research perspectives using advanced geophysical techniques, as detailed in recent studies (Khatereh et al., 2022; Zuo et al., 2021, 2024).

Lunar crust magnetic anomaly vector data, captured at an altitude of 30 km and derived from a magnetic field model (Wieczorek, 2018), are integrated into a 3D lunar model using a lunar-centric global Cartesian coordinate system (Fig. 1a & 1b), rather than traditional local 2D map projections. This innovative 3D vector model offers a distinctive perspective on lunar magnetic anomalies, illustrating the direction and intensity of crust magnetization anomalies through an intuitive visualization. To address the challenge of displaying only a limited portion of the data effectively, we create Movie S1—an animation that depicts a 360° rotation of the 3D lunar model, ensuring comprehensive visualization of the data.

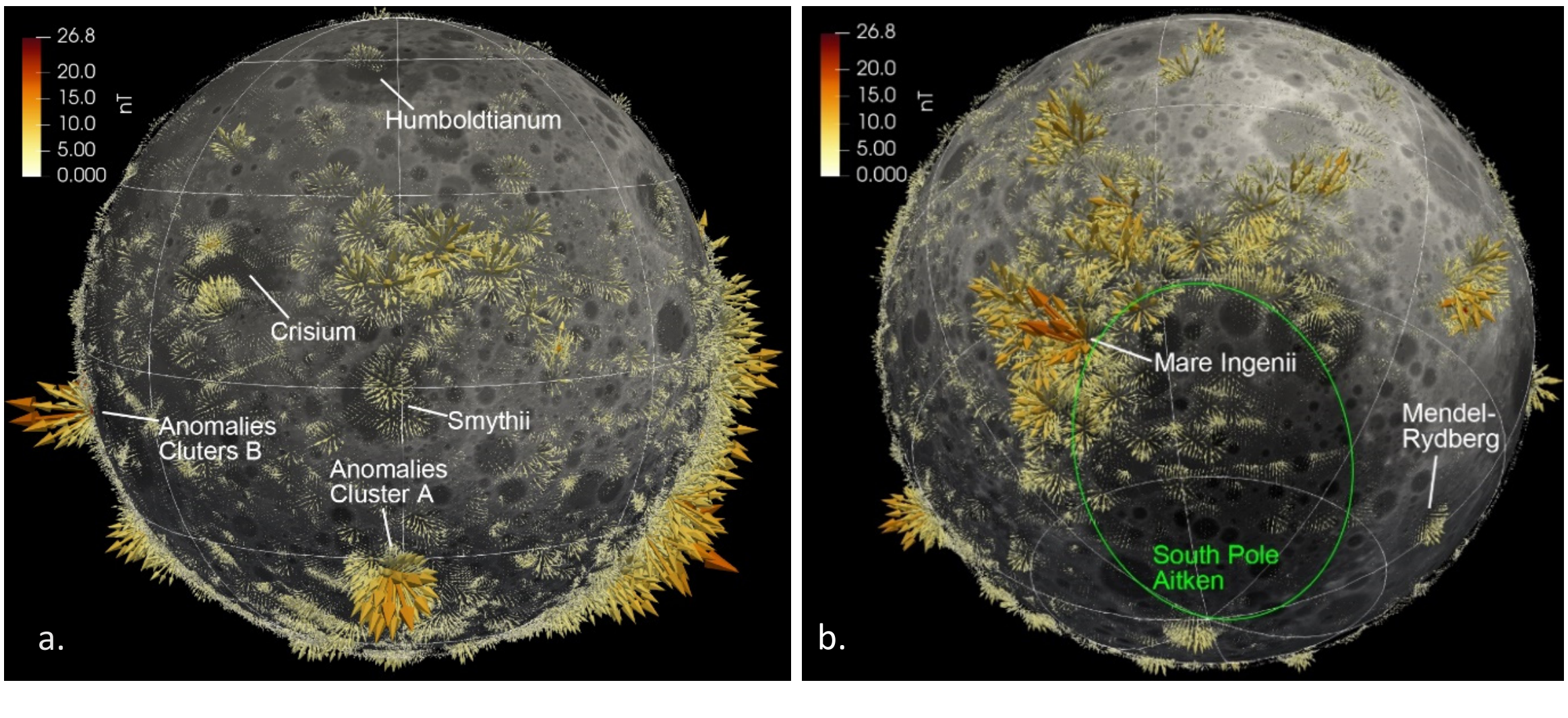

**Fig. 1: 3-D Model of Lunar Crust Magnetic Anomalies Vectors map.**

Homogeneously sampled at a 30 km orbital altitude, panel (a) displays the near side and panel (b) the far side of the Moon.

New features of lunar magnetization anomalies are elucidated by the 3D model. It is now determined that the crustal anomalies are not magnetized by an ancient core field; otherwise, a global magnetic field pattern akin to Earth's geomagnetic field would be evident (Fig.S1). The 3D model reveals that lunar magnetic anomalies comprise numerous regional clusters (Fig. 1a & 1b), each exhibiting characteristics of a magnetic dipole field. For instance, central anomalies are observed in the Smythii and Humboldtianum basins, with intense anomalies located in highland regions (clusters A & B) and maria (Fig. 1a). Additionally, these intense clusters are concentrated near the edge of the South Pole-Aitken basin on the far side (Fig. 1b). These dipole-like clusters typically align upward or downward, perpendicular to the lunar surface, irrespective of latitude. Only a few anomaly clusters exhibit different directions, likely influenced by subsequent impact events. This pattern indicates that the lunar crust magnetic anomalies are likely magnetized by local magnetic fields, not by the ancient core field.

Secondly, the nature of the dipole-like anomaly clusters suggests that the magnetizing fields are not transient phenomena generated solely during impacts. While basins such as Humboldtianum and Smythii exhibit pronounced anomalies at their centers (Fig. 2a), most lunar impact basins do not feature a regular central anomaly (Fig. 1a & 1b). Numerous intense anomaly clusters are located in the highlands or maria, rather than within the basins themselves—such as clusters A and B in Fig. 1a, and the anomalies along the edge of the South Pole-Aitken basin (e.g., Ingenii Mare in Fig. 2b). If transient fields from impacts were the primary local magnetizing force, the magnetization anomalies recorded would display a closer relationship with basin structures. However, this is not observed in the distributions of anomalies.

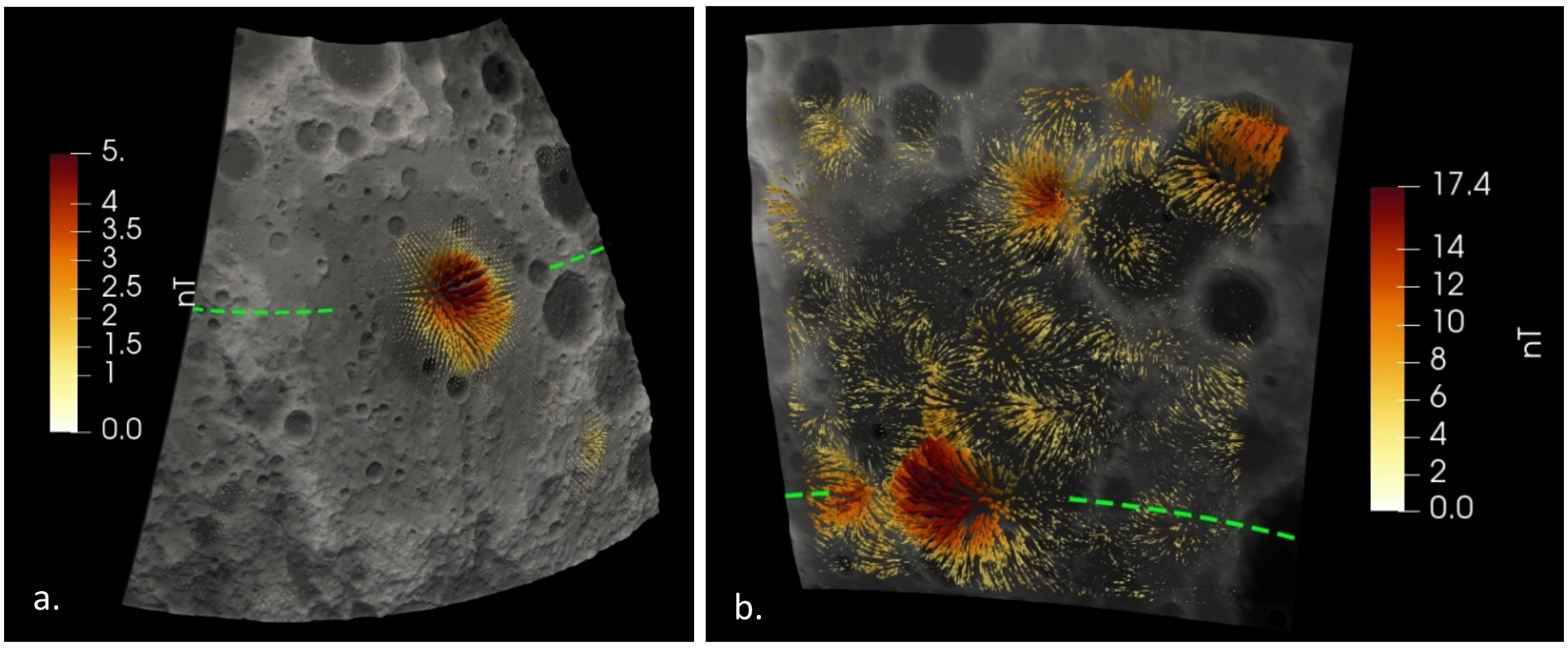

**Fig. 2: Magnetic Anomalies in Lunar Basins at 30 km Altitude.**

Panel (a) shows the Mendel-Rydberg basin; panel (b) the Ingenii mare (basin). Green dashed lines mark the sections for subsequent figures.

To validate our findings, we invert the subsurface magnetization models of the Mendel-Rydberg basin, characterized by a regular magnetic anomaly at its center (Fig. 2a), and Ingenii Mare, marked by irregular anomaly clusters (Fig. 2b). This analysis aims to explore the signatures of magnetizing fields, reconstructed by the inverted magnetization models (Zuo et al., 2024), as shown in Figs. 3a and 3b. Clear animated versions are available in the supporting materials, Movies S2 and S3.

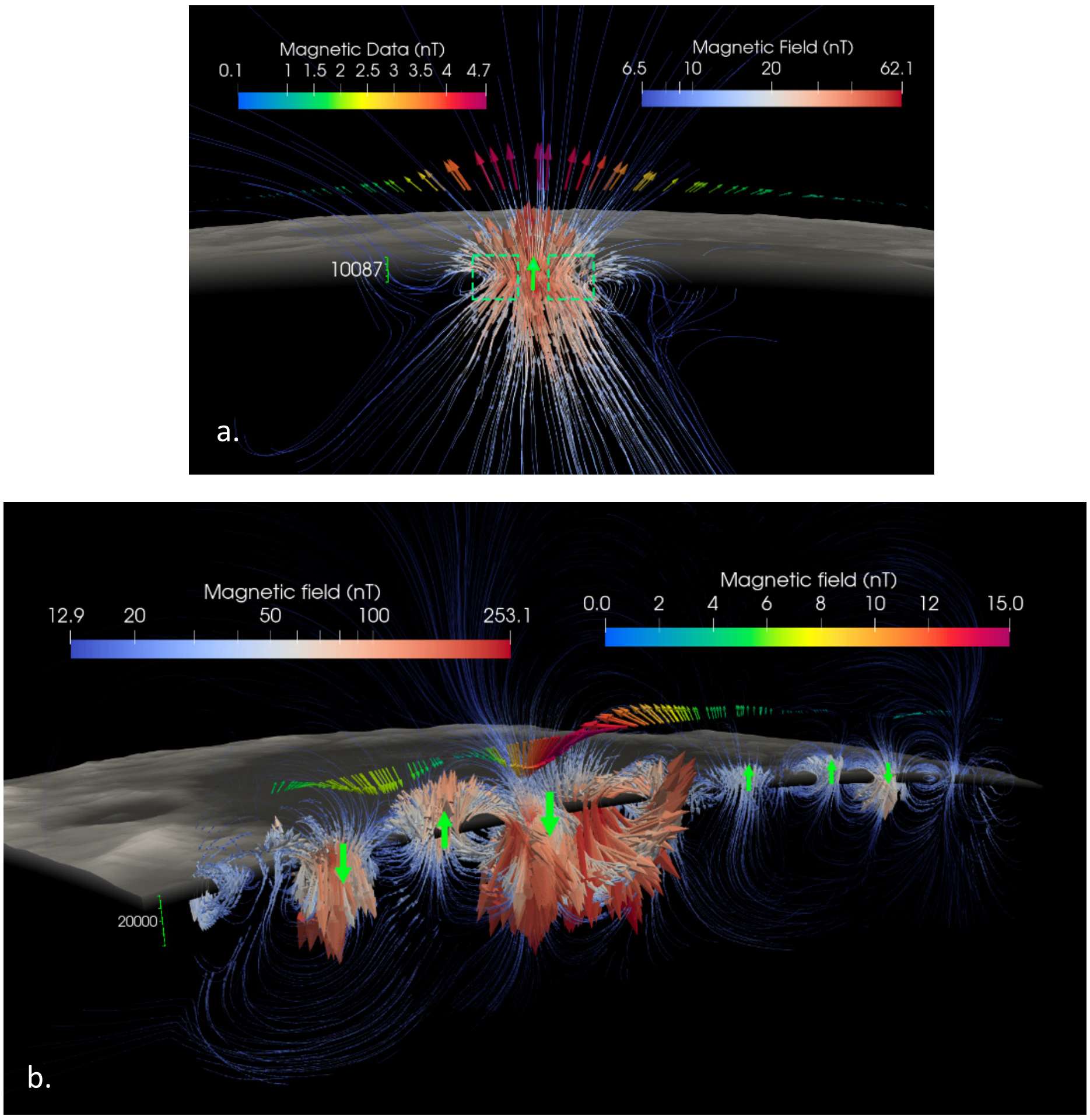

**Fig. 3: Sections of Inverted Magnetization 3-D Model.**

Panel (a) depicts Mendel-Rydberg and panel (b) Ingenii Mare. Arrows above the crust indicate orbital data (length corresponds to strength); arrows under and near the crust show the magnetization model for (a) and the recovered magnetic field for (b). Green arrows identify potential dipole-shaped sources. The slice positions are indicated by green lines in Fig.2a & 2b.

As depicted in Fig. 3a, the crust of the Mendel-Rydberg basin is influenced by an isolated, upward dipole-like magnetizing field. The potential source position of this field, indicated by a green arrow, is located in the shallow crust. The 3D magnetization model for Ingenii Mare reveals several weak dipole-like fields at the basin edges, marked with green arrows, and a central, intense toroidal electrical current source generating a strong magnetizing field. It is well-established that a magnetic dipole is analogous to a toroidal current with zero radius, and broad, intense dipole-like fields are typically produced by electrical current loops. This toroidal current induces a poloidal magnetic field, with lines circulating around the current source as illustrated in Fig.S2 and S3. We invert the magnetization model for Ingenii Mare and reconstruct the map of ancient electrical currents (Fig. 4a) using the equivalent electrical current method (detailed in supporting online material Text S1).

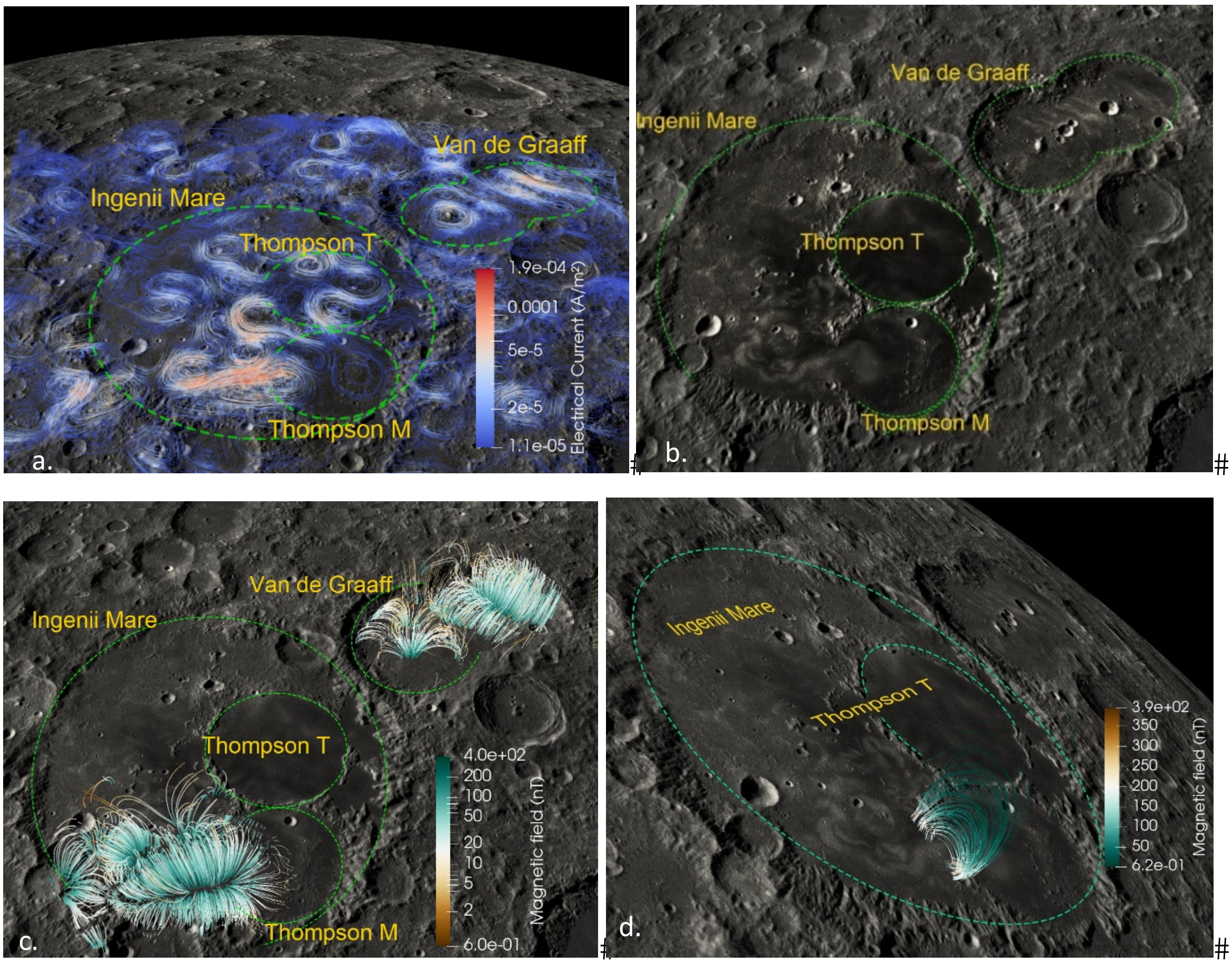

**Fig. 4: Recovered Electric Currents and Related Magnetizing Field in Ingenii Mare Region.**

(a) Map of electrical currents. Animated version in supporting online material Movie S4 & S5. (b) Swirls in Ingenii Mare and Thompson basins. (c) Recovered 3-D magnetic field in Thompson M and Van de Graaff basins, omitting other regions. (d) Swirl S and a slice of its associated 3-D poloidal magnetized field.

The morphology of the electrical currents depicted in Fig. 4a closely mirrors the lunar swirls observed in Ingenii Mare, as shown in Fig. 4b. Notably, the sinuous S-shaped current traversing the rim of the Thompson basin and the linear current in the Van de Graaff basin are vividly illustrated. The 3D magnetic field proximal to these swirls is constructed using the inverted magnetization model, as displayed in Fig. 4c, with a cross-section of this field overlaid with the S-shaped swirl in Fig. 4d. Application of Ampère's law confirms that the morphology of the swirls aligns with the distribution of electrical currents, which in turn generate the surrounding poloidal magnetic field, as detailed in Figs. 4c and 4d. While these magnetizing fields are estimated using the magnetization model, it is important to note that only the distribution of the ancient magnetizing fields is recovered. The intensity of these fields will be addressed separately in a subsequent section.

In this study, we hypothesize that lunar swirls are remnants of electrical currents that flowed through the lunar crust during periods of ancient dynamo activity. These currents attracted plagioclase-rich dust via electric fields, leading to the formation of high-albedo optical swirls. Furthermore, these electrical currents generated intense magnetic fields that subsequently magnetized the lunar crust, resulting in the magnetization anomalies currently observed by orbiters. A detailed description of the magnetizing procedures is provided in the supporting online material (Text S2).

We argue that electrical currents, rather than magnetic field shielding or intrusive magmatic dikes, are more likely responsible for forming lunar swirls. Firstly, electrical currents can generate intricate structures when flowing through regions of high electrical conductivity, such as intrusive tubes or dikes. Although our inverted electrical current models (Fig. 4a) do not exhibit the same fine structure as the swirls (Fig. 4b), this discrepancy is likely due to limitations in the resolution of orbital data. In contrast, the 3D magnetic field, which spreads over a wide area (Fig. 4c), is unlikely to create such detailed structures through shielding effects alone. Furthermore, analogous high-albedo branch anomalies, formed by ejected plagioclase-rich dust from meteoroid impacts (e.g., in the Oceanus Procellarum region, Fig.S4), are well-preserved without the influence of magnetic field shielding.

The most direct evidence for determining the origin of the lunar swirls—whether from electric currents or magnetic fields—lies in their alignment with the corresponding field structures. If the boundaries of the swirls align with the vertical edges of the magnetic field, it would suggest that magnetic shielding is responsible for their formation. Conversely, if the swirl boundaries are situated at the center of the 3D magnetic field (indicative of potential poloidal current locations), it would imply that the swirls are formed by the accretion of plagioclase-rich dust due to poloidal electric currents. Analysis of the 3D magnetic field slice (Fig. 4d) reveals that the swirls are located at the positions of potential poloidal electric currents, rather than at the vertical boundaries of the magnetic field. This indicates that the swirls are formed by poloidal electric current accretion rather than magnetic shielding.

Additionally, the formation of swirls due to magma intruding and flowing on the lunar surface is also unlikely. For instance, the fine striped shape of the swirls in the Van de Graaff basin (Fig. 4b) does not match the expected patterns from surface-flowing magma. If the swirls were formed by such magma, the magnetizing field would have to be either the core field or a field generated by conductive fluid flow within the core field, following magnetohydrodynamic (MHD) processes. However, neither scenario aligns with the observed magnetic data in terms of field direction and intensity.

Rein-Gamma is another prominent large lunar swirl complex located in the Oceanus Procellarum region. Intense magnetic anomalies have been observed above the narrow, striped, high-albedo swirl region. As illustrated in the magnetic field data slice (Fig. 5a), the strongest magnetic anomalies coincide with the brightest part of the swirl, where the magnetic field direction changes rapidly. The relationship between these swirl complexes and the magnetic anomalies remains unresolved. To investigate this, we invert the subsurface magnetization model and recover the distribution of the ancient electric currents, as shown in Fig. 5b.

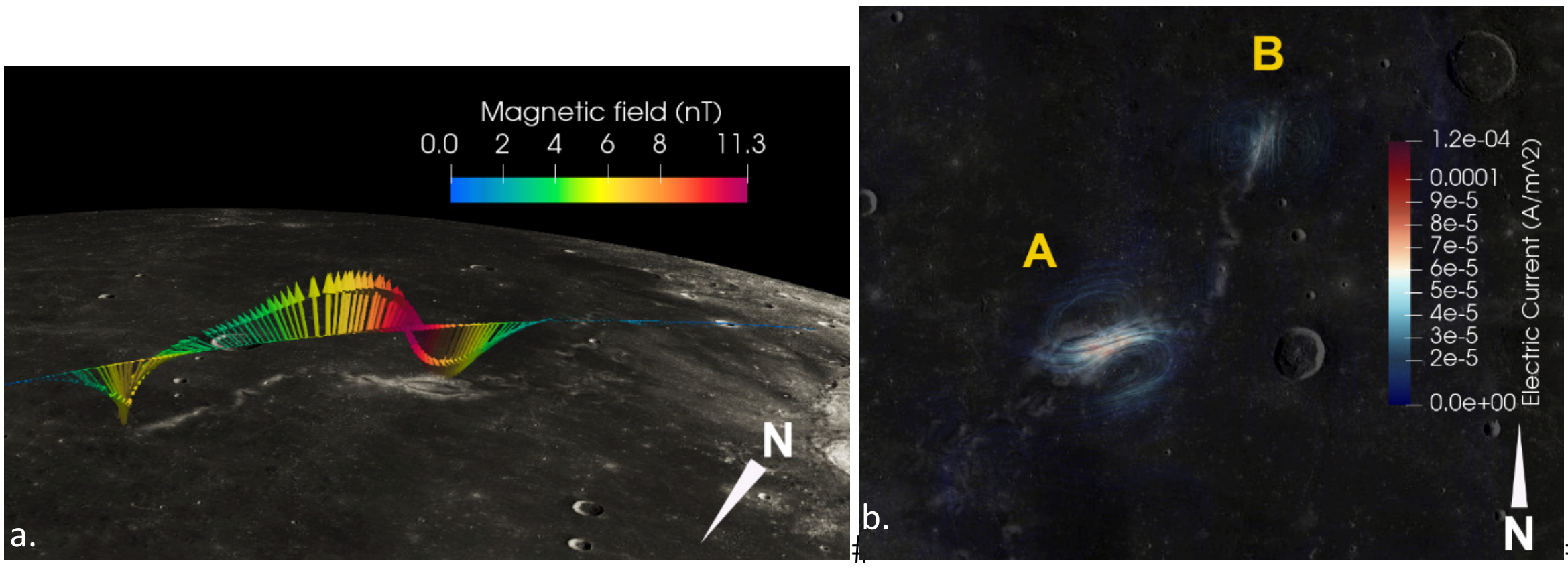

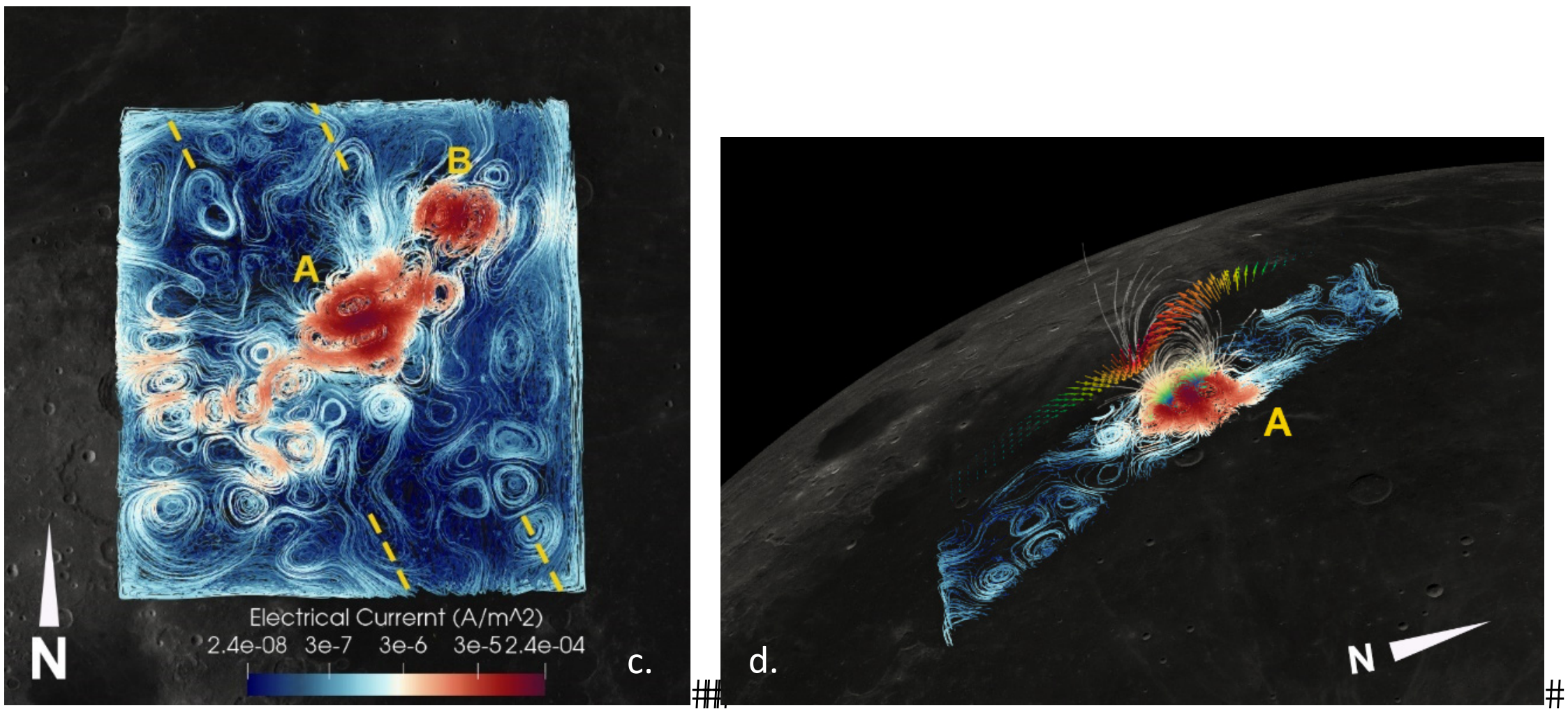

**Fig. 5: Reiner-Gamma Swirls and the Estimated Electric Currents.**

(a) Slice of magnetic observed data above swirls at 30 km altitude. (Slice position shows in Fig.S5) (b) Estimated electric current circulations depicted with a normal color scale. (c) Complex electric currents painted with a logarithmic color scale. Dashed lines mark the section positions in panel d. Animated version in supporting online material Movie S6. (d) Sections of the recovered 3-D poloidal field (yellow line) and the orbital observed data vectors (colored arrows). Letter A marks the primary central electric currents, and B marks the smaller one.

The morphology of the inverted electrical currents closely aligns with the distribution of the Reiner-Gamma swirls (Fig. 5b), suggesting that intense electrical currents once flowed through the lunar crust, accumulating plagioclase-rich dust and forming the swirls. To clearly depict the intricacies of electrical current circulation, a logarithmic color scale is employed (Fig. 5c), revealing a continuous, sinuous, and complete electric current circuit. These electrical currents generate a 3D poloidal magnetic field (Fig. 5d), which magnetizing the lunar crust and corresponds with the orbital magnetic anomalies observed, as illustrated by color arrows in Fig. 5d.

Notably, the recovered currents in Ingenii Mare (Fig. 4a) appear much rougher compared to those in Reiner-Gamma (Fig. 5c). This disparity is attributed to the Imbrium impact event, which likely disrupted the continuity of the remanent magnetization in Ingenii Mare, dating back to the Pre-Nectarian epoch. In contrast, the Oceanus Procellarum region, having experienced fewer impact events, has preserved the electric current magnetization information more completely.

To explore the origins of these electric currents in the lunar crust, we invert the 3D density and magnetization subsurface structures of the relevant regions. The Bouguer gravity anomaly data, essential for our analysis, are derived from the GRGM1200B RM1 model (Goossens et al., 2019), which assumes a constant crustal density of 2500 kg/m³. Vertical sections beneath the swirls are carefully selected from the inverted 3D density model to further elucidate the underlying structures.

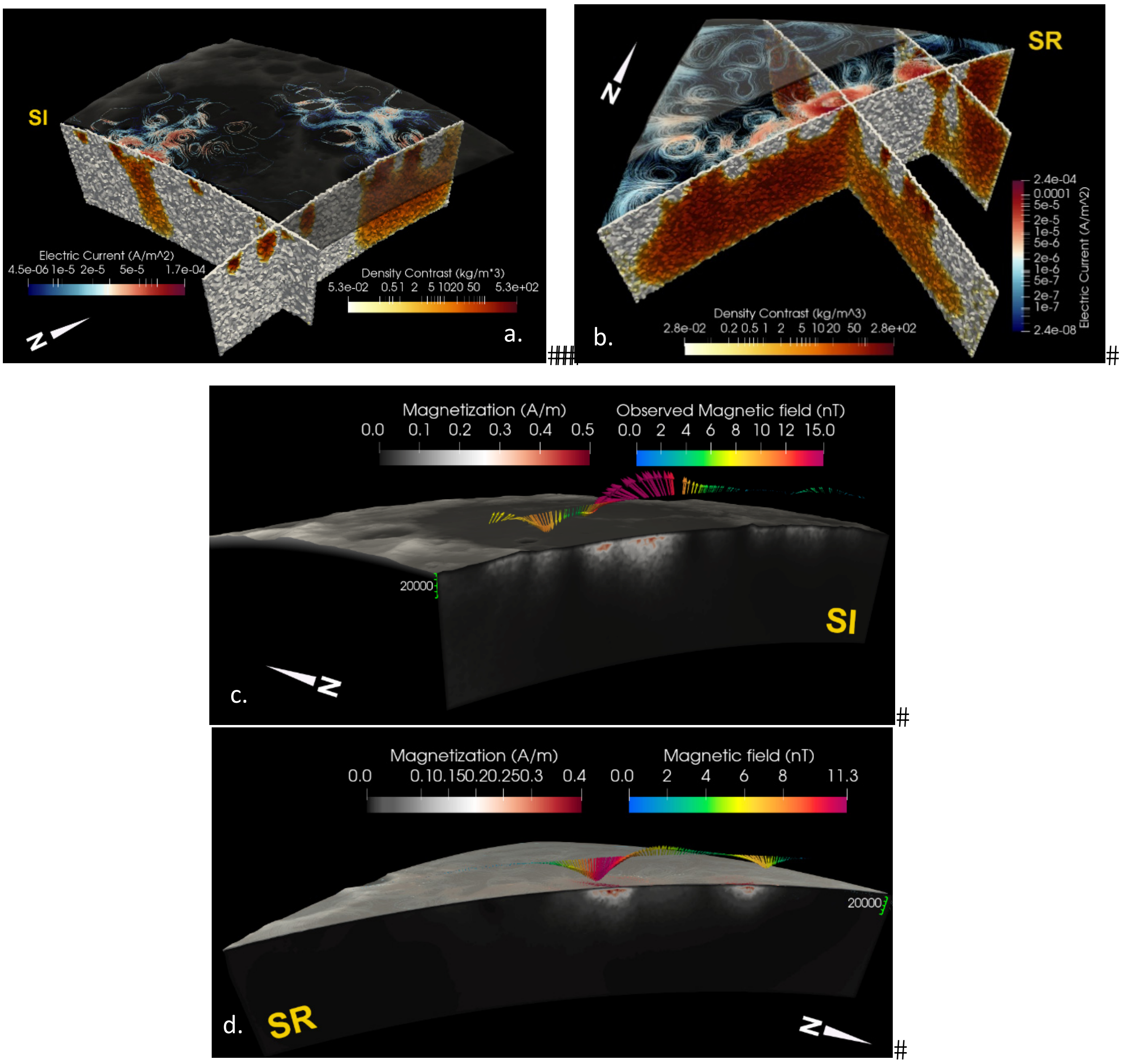

**Fig.6 Inverted density model and Magnetization model in Ingenii regions and Reiner-Gamma.**

(a) Density section of Ingenii Mare ( [-290, 390] (kg/m$^3$)]. Animated version in supporting online material Movie S7. (b) Magnetization section of Ingenii Mare. (c) Density section Reiner-Gamma ( [-108, 157] (kg/m$^3$)] ). Depths of the sections are all 150 km. (d) Magnetization section of Reiner-Gamma. Positions of SI and SR are marked in Fig.S6a & 6b.

Beneath the expansive S-shaped swirl in Ingenii Mare, a high-density intrusion extends from the deep lunar mantle to the crust, bifurcating into two branches near the lunar surface (Fig. 6a, section SI). Similarly, a substantial high-density rock intrusion is observed beneath the stride swirl in the Van de Graaff basin (Fig. 6a, another section). In the Reiner-Gamma region, two high-density intrusive tubes emerge from the deep mantle beneath the prominent swirls (Fig. 6b, section SR).

The inverted magnetization model for Ingenii Mare indicates that the magnetization strength reaches 0.5 A/m beneath the S-shaped swirls (Fig. 6a). In the Reiner-Gamma region, a magnetization strength of 0.4 A/m is observed beneath the swirls (Fig. 6b). Considering the diffusion effect inherent in the inversion procedure, the actual magnetization strength may be higher than these values. For the 0.4 A/m magnetization in Reiner-Gamma, the corresponding magnetizing field strength should be approximately 318 μT (thermoremanence susceptibility is set to $1.58 \times 10^{-3}$ SI as mare basalts, Wieczorek et al., 2012). This value significantly exceeds the strength of the core field (40μT - 100 μT).

Here, we propose that intense ancient electrical currents originated from the deep interior of the Moon. Although the movement of lava flows within the ancient lunar magnetic field can generate electrical currents, their density is insufficient. For instance, consider an upwelling lava flow beneath the crust in the Reiner-Gamma region, characterized by high conductivity of 100 S/m and a velocity of 10 m/s. The induced current, as it traverses a magnetic field of 100 μT, would only produce a magnetic field of 3 μT, significantly lower than the observed magnetizing field strength of 318 μT (detailed in the supporting online material, Text S3). Furthermore, even if drastic changes in the lunar core's magnetic field significantly increased the velocity parameters, leading to stronger electrical currents, such a scenario would still not align with the observed patterns of intrusive rocks (Fig. 6a & 6b). The most plausible source of these currents is the powerful electrical activity generated by ancient core dynamo processes. Under the thermal influence of the core, high-density, high-conductivity mafic lava ascends to the crust-mantle boundary. The intense currents produced by lunar dynamo activity are partially channeled through these pathways to the lunar crust. Once at the surface, these currents flow within the crust, forming the swirls and generating the observed magnetic fields.

A 10 km horizontal electrical current with a density of 13 A/m², confined within a cross-sectional area of 3 km by 1 km, is introduced at the crust-mantle boundary at a depth of 10 km and flows horizontally along the crust (Fig.S7). In this region, the crust's conductivity is set according to the inverted density model, with values of 0.1 S/m for densities greater than 2500 kg/m³ and 0.01 S/m for densities smaller than 2500 kg/m³. The electrical current diffuses upward from the base of the crust to the surface. Fig. 7a shows that the distribution of electrical current on the crust's surface exhibits an elliptical shape, resembling the morphology of Swirl A in Reiner-Gamma (Fig. 5a). This electrical current generates the magnetizing field, which in turn magnetizes the rocks. The magnitude of it are approximately three orders of magnitude higher than the orbital data, recording 22.7 μT at orbital altitude and 469 μT on the lunar surface (Fig.S8, Movie S8). The rocks have a thermoremanent susceptibility set at $1.58 \times 10^{-3}$ SI (mare basalt). The resulting magnetic field anomaly, simulated based on the magnetization model of these rocks, is presented in Fig. 7b.

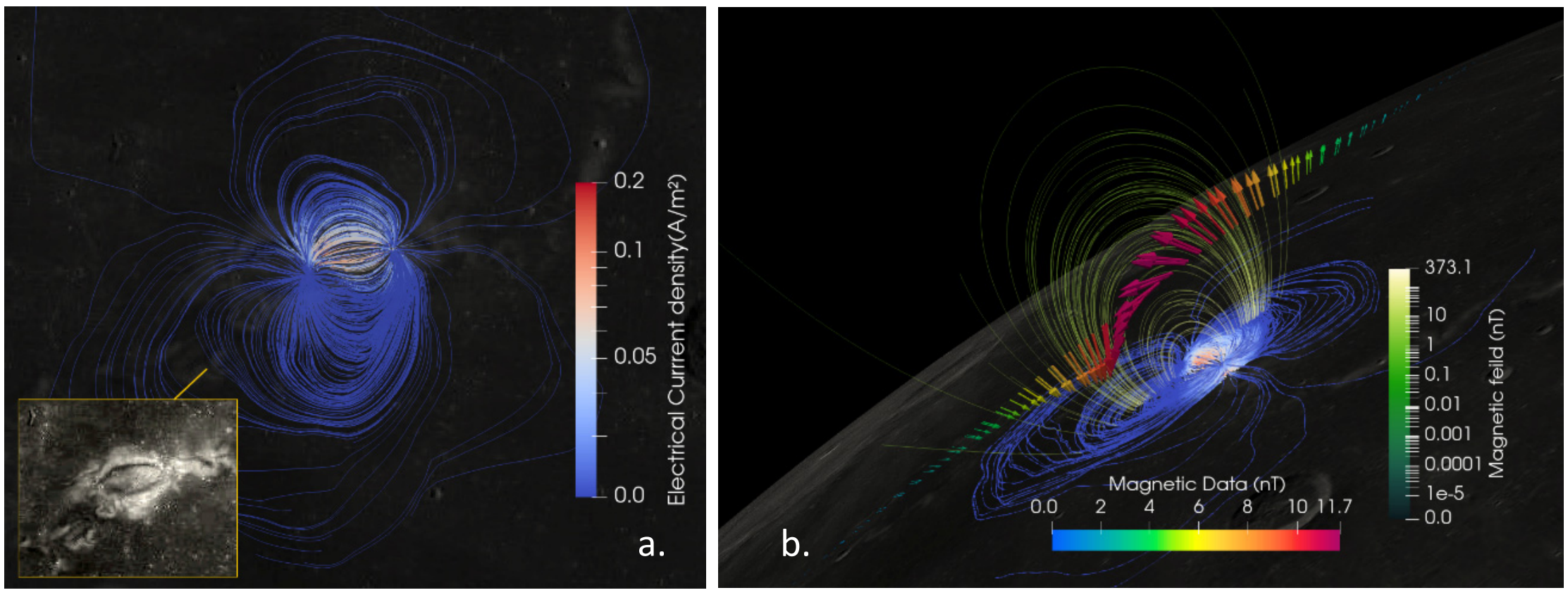

**Fig. 7. Synthetic Simulation Experiment.**

(a) Distribution of electric currents on the crust surface. (b) Resultant magnetic anomalies produced by the magnetized rock, illustrated with color arrows at orbital altitude. Green lines depict a slice of the 3-D magnetic field near the surface. Animated version in supporting online material Movie S8.

The strength and distribution of the simulated rock remanent magnetization anomaly at an altitude of 30 km (Fig. 7b) is 11.7 nT, closely aligning with the actual orbital observation data (11 nT, Fig. 5a). At the lunar surface, the magnetic anomaly reaches 373 nT. This simulation effectively demonstrates that deep interior electrical current sources can generate electric and magnetic fields that closely match the distribution of lunar swirls and orbital magnetic anomalies.

In addition to anomalies in the mare regions, our analysis encompasses three types of mascon basins. Notably, in the Mendel-Rydberg basin, an isolated anomalous cluster (Fig. 2a) is closely linked with the basin's structure. The density inversion model suggests that this density distribution stems from an impact event. It reveals a bowl-shaped mascon with a 170 km radius, encircled by a negative density ring structure with a 340 km radius, exhibiting a density contrast ranging from -316 to -80 kg/m³ (Fig. 8a). The model slice illustrates high-density materials uplifting at the shallow rim, potentially forming a conduit for electrical current conduction that connects to the deep lunar interior (Fig. 8b). The inversion of ancient electrical currents and the associated intense orbital magnetic anomalies, positioned above the rim of the mascon (Fig. 8c & 8d), corroborate this interpretation.

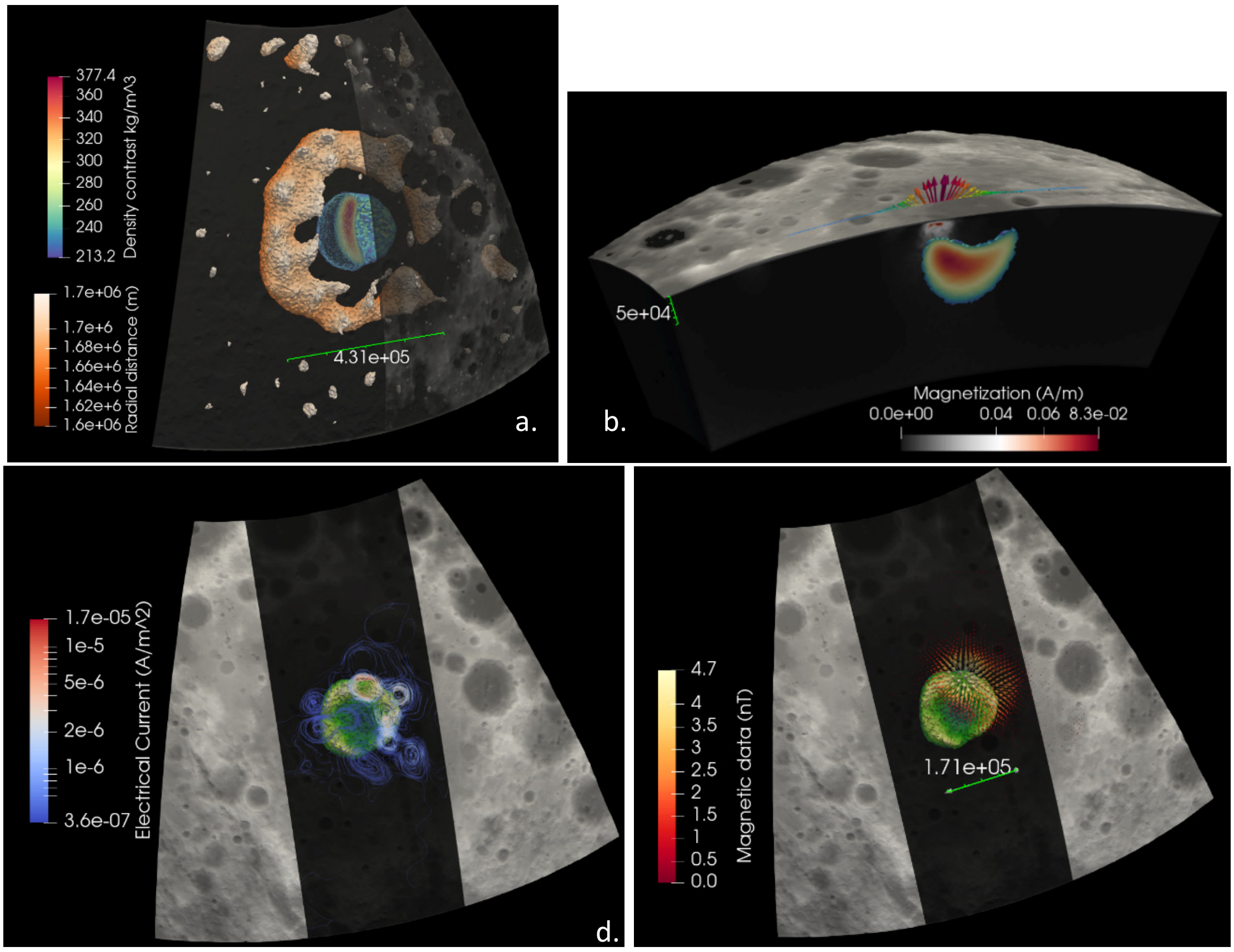

**Fig. 8: Density, Electrical Current, and Magnetic Anomalies Model in the Mendel-Rydberg Basin.**

(a) Density model beneath the lunar crust. Animated version in supporting online material Movie S9. (b) Cross-section of the near-surface magnetization model juxtaposed with the deep mascon density model. (c) Electric eddies depicted above the mascon, with color indicating depth. . Animated version in supporting online material Movie S10. (d) Vectors of orbital observed data over the mascon.

Crisium Mare, a large basin with a diameter of 555 kilometers, features a substantial ring-shaped mascon with a radius of 330 km, spanning depths from 25 to 120 km (Fig. 9a). This model further illustrates a rim uplift structure. Notably, the depth of magnetization in Crisium Mare is significantly greater, extending below 60 km (Fig. 9b), in contrast to Mendel-Rydberg, where it is less than 10 km. The magnetization in Crisium Mare aligns precisely with the uplifted rim of the mascon (Fig. 9b), corroborating the patterns of inverted ancient currents (Fig. 9c) and aligning with observed data (Fig. 9d). We hypothesize that an impact event catalyzed the upwelling of high-conductivity materials from the deep lunar interior, triggering intense electrical currents near the crust that are responsible for the anomalies observed in the Crisium basin.

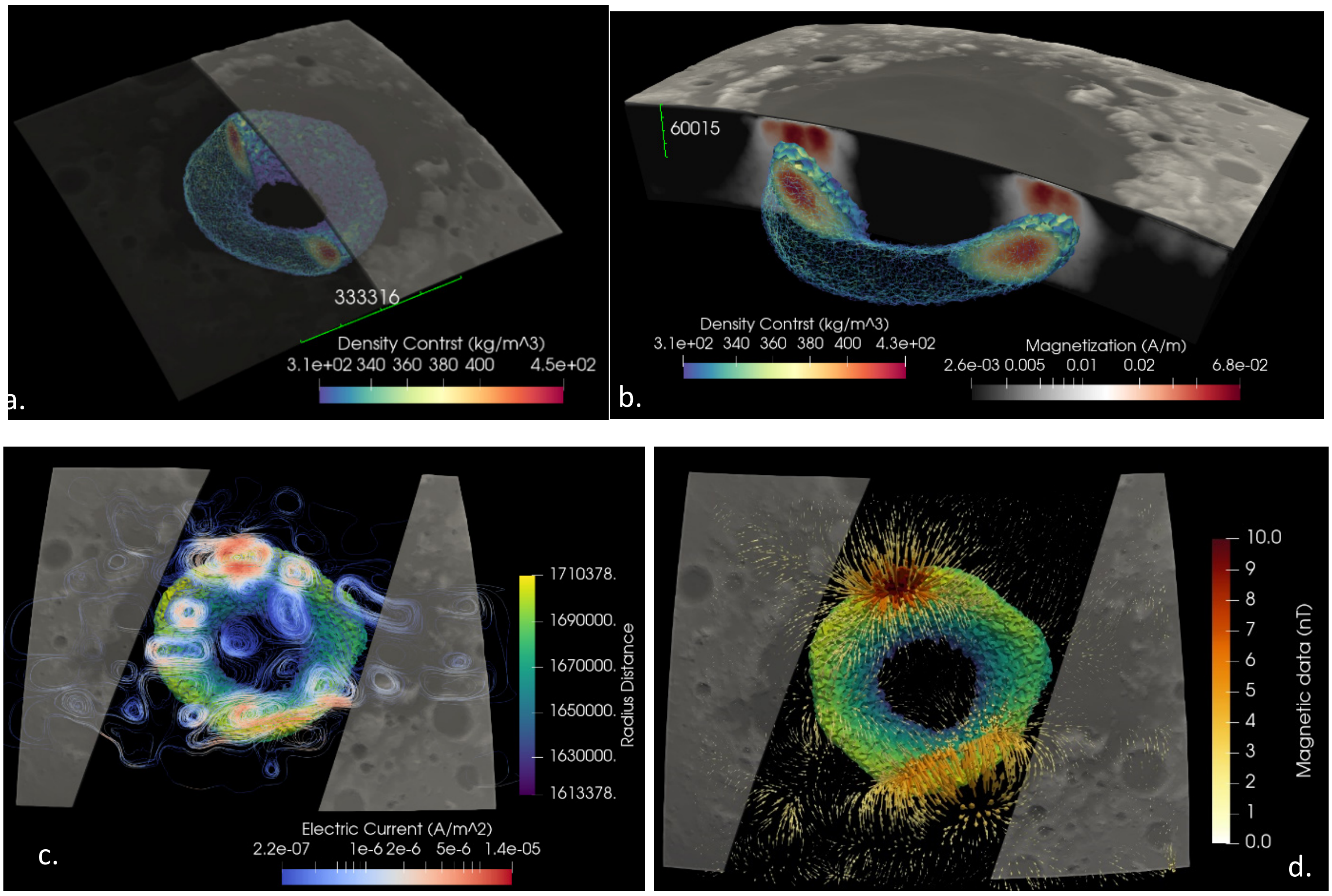

**Fig. 9: Density, Electrical Current, and Magnetic Anomalies Model in Crisium Mare.**

(a) Density model beneath the lunar crust. (b) Cross-section of the magnetization model integrated with mascon. (c) Electric eddies depicted above the mascon, with colors representing depth. (d) Vectors of orbital observed data over the mascon.

Although large impact basins with mascons, such as Crisium, typically do not exhibit significant swirls, intense electrical currents have been identified in the Crisium basin (Fig. 9c). To further investigate, we enhanced the optical image of the basin's central region (Fig.S9), which revealed weak swirl anomalies corresponding to the locations of the inverted currents. This finding supports the presence of weak electrical currents at the surface, conducted from the shallow rim of the mascon.

As depicted in Fig. 9b, the magnetized rock layer in this region reaches a thickness of 30 km. These basalts required at least 28.5 million years to cool from their molten state to their Curie point (600°C). The magnetizing field must have formed either subsequently or been maintained throughout this period, suggesting that the source of these electrical currents is endogenous, originating from deep within the Moon's interior, rather than being a direct result of an impact-generated ambient field.

The Leibnitz basin, a unique lunar crater, is distinguished not only by its magnetic anomalies and a central mascon but also by the presence of swirls at its center (Fig. 10a), making it an exemplary case for exploration. We have mapped the electrical currents (Fig. 10b), finding that their distribution closely aligns with the swirl patterns observed in the basin. The source of magnetization in the Leibnitz basin is located at a depth within 10 km, akin to that in the Mendel-Rydberg basin (Fig. 10c). However, the mascon here is shallower compared to those in other basins (Figs. 8b & 9b), with its upper part extending very close to the lunar surface (<5 km). The apex of the mascon is not directly at the center of the current loop but is positioned just below the eastern part of the central current loop. This configuration correlates with the pattern of strong upward currents on the lunar crust surface. Unlike the deeper mascon in the Crisium basins, the high-conductivity mascon in the Leibnitz basin is comparatively shallow, facilitating the movement of powerful currents from the deeper lunar interior to the lunar crust. This process leads to the accretion of plagioclase-rich dust and the formation of distinctive lunar swirl anomalies on the surface. These dynamics suggest that the shallow mascon not only supports but enhances the electrical activity that shapes these surface features.

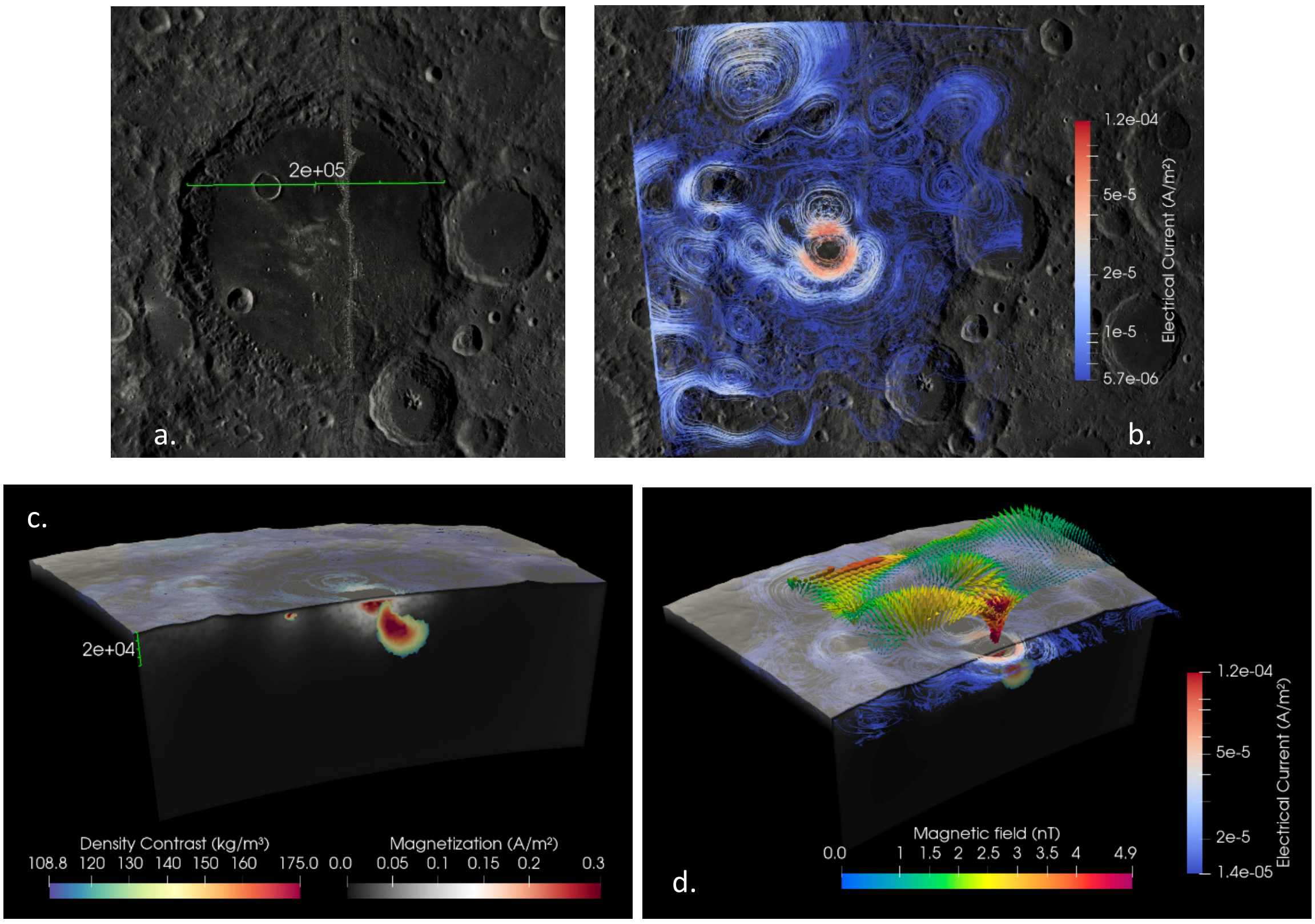

**Fig. 10: Density, Electrical Current, and Magnetization Model in Leibnitz Basin.**

(a) Swirl anomalies in the Leibnitz basin; (b) Inverted electrical eddies; (c) Combined magnetization and density models; (d) Correlation between magnetic anomalies and the current source. . Animated version in supporting online material Movie S11.

The depth information of the magnetization field provides crucial insights into the origin of the magnetizing forces affecting the lunar crust, revealing whether they stem from extralunar sources such as solar wind, plasma, or from endogenous sources within the Moon itself. To further investigate, we constructed 3D magnetic field models using inverted magnetization data from the Mendel-Rydberg basin, Crisium basin, and Ingenii Mare.

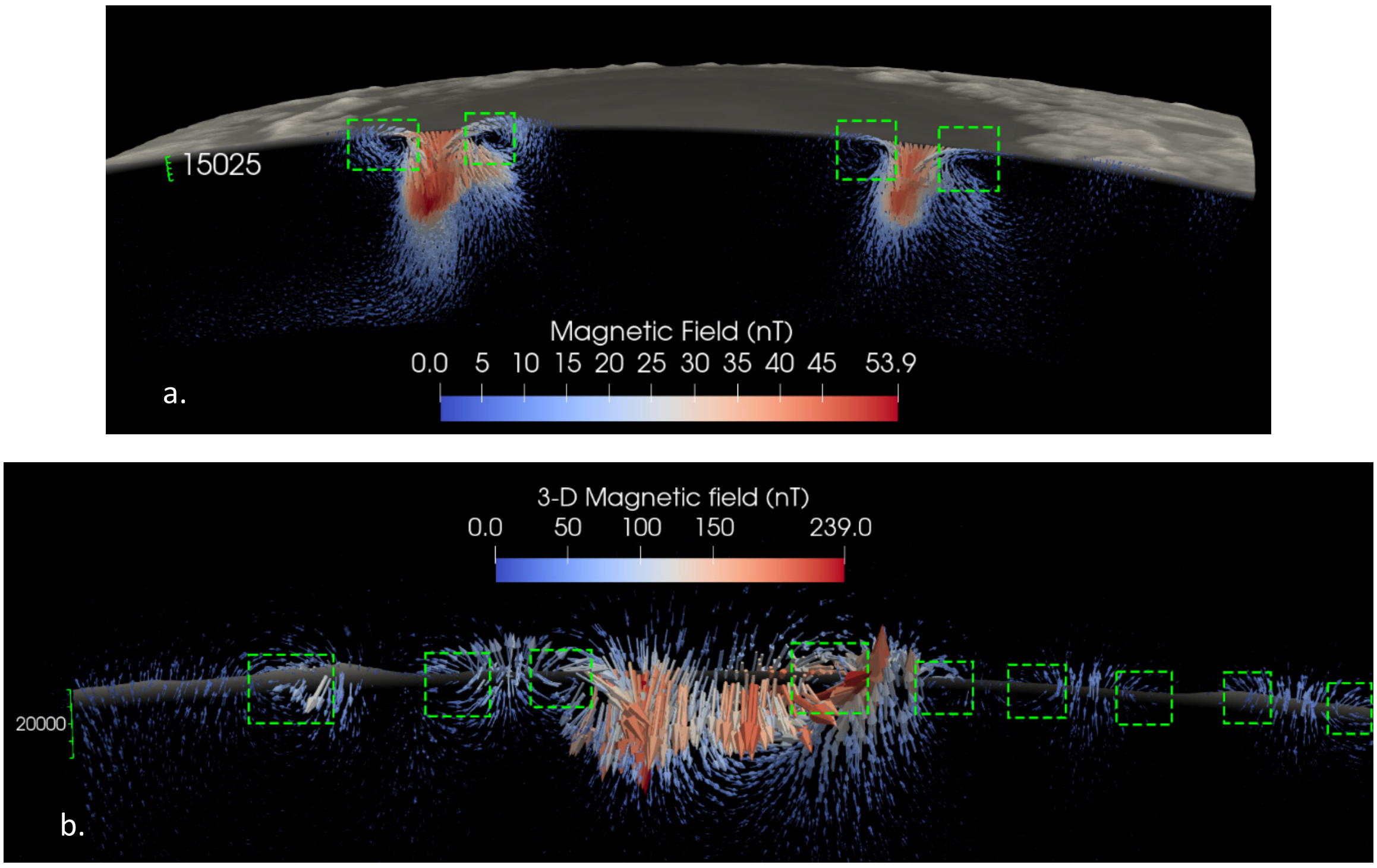

**Fig.11 Depth of the electrical current source.**

(a) Crisium basin. (b) Ingenii Mare. Positions of the sections as same as above cases. Green dash rectangles mark the flipping positions of magnetic fields. Only the fields below of lunar surface are draw for Fig.10a & 10b.

In a typical vertical magnetic dipole field, a reversal in direction is typically observed at the depth of the source (Fig. S2). This reversal feature, identified at a depth of 10 km within the Mendel-Rydberg basin, accurately locates the source of the magnetizing field within the crust (Fig. 3a). Similarly, a reversal is observed in the Crisium basin model (Fig. 11a) at a depth within 15 km. In Ingenii Mare (Fig. 11b), the presence of both dipole and loop sources is evident, with all reversal characteristics indicating that the source is located within about 10 km of the crust. These cases collectively suggest that the ancient magnetizing fields are generated by endogenous sources within the lunar crust, rather than from extralunar influences.

The magnetizing field generated by a current exhibits a characteristic non-uniform distribution, with field strength diminishing rapidly as distance from the source increases (Fig.S10). For instance, in the synthetic example, the field strength peaks at 2 mT near the current source. By utilizing the inverted magnetization models of the discussed regions and referencing the estimated range of thermoremanent susceptibility for lunar rocks (Wieczorek, 2012) along with Equation S7, we estimate the maximum possible strength of the magnetizing fields in these study areas, as detailed in Fig. 12.

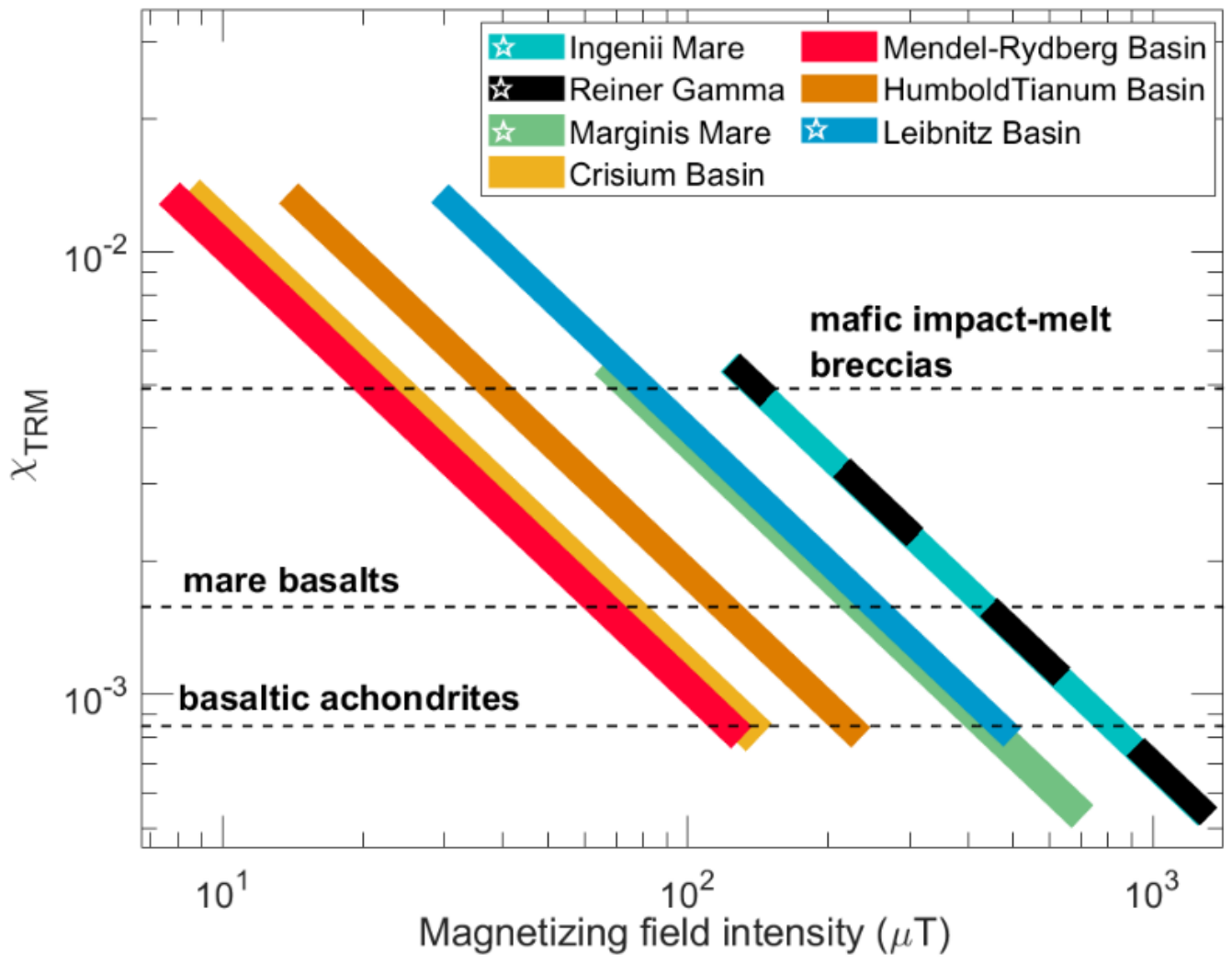

**Figure. 12 Strengthes of the ancient magnetizing field of study region in lunar.**

☆ marks the region have swirl anomaly. Dash line used in rep Table S1.

The intensity of the paleomagnetic field in regions with lunar swirls is significantly greater than in regions devoid of such features. For instance, as depicted in Fig. 12, in areas primarily composed of mare basalts like the Humboldtianum basin, the magnetizing field strength is approximately 100 μT. In contrast, this strength escalates to 400 μT in the Reiner-Gamma and Ingenii Mare regions. If these regions predominantly contain basaltic achondrites, the ancient magnetizing field could reach up to 1000 μT.

The significant strengths of magnetizing fields suggest that magnetized materials may include not only extralunar projectile materials with high thermoremanent susceptibility but also indigenous lunar rocks, such as mare basalts, pristine highland rocks, and other native lunar formations. The presence of swirls highlights regions where ancient, intense electrical currents once traversed the lunar crust, unlike areas without swirls, which lack the strong surface electrical currents necessary to generate such optical anomalies. This distinction elucidates the consistent association of strong magnetic anomalies with lunar swirls and underscores the presence of isolated intense magnetic anomalies.

Due to the absence of a dense atmosphere and other active layers on the lunar surface, swirls have been well recorded and preserved. In contrast, Earth's surface is significantly influenced by its atmosphere, water, and biosphere layers. If electrical currents were to conduct to Earth's surface, it is unlikely to produce and preserve such swirl anomaly. Currently, electrical discharges have been observed during volcanic eruptions, which may be contributed by this mechanism. Inspired by the phenomenon of lunar swirls, it can be postulated that celestial bodies with core dynamo activity may exhibit surface discharge phenomena. Our study may aid the further exploration and observation of Earth and extraterrestrial bodies.

## Data availability

The magnetic and gravity dataset in this study is available at

https://doi.org/10.6084/m9.figshare.26465086.

## Code availability

The magnetization current estimate code is available at

https://figshare.com/articles/dataset/code/22585969?file=48171805

## Acknowledgements

## Author contributions

**B.X.** designed the project, performed the experiments and discussion, and wrote the manuscript. X.H. contributed to the funding. Z.W. contributed to the funding. Y.C. contributed to data analysis. M.A.K. contributed to methods conception.

## Competing interests

The authors have no conflicts of interest to declare.

## Additional information

Supplementary Information.

Correspondence and requests for materials should be addressed to B.X.

Supplementary Materials for

# Lunar Swirls Unveil the Origin of the Moon Magnetic Field

Boxin Zuo*[1], Lizhe Wang[1] , Xiangyun Hu[2], Yi Cai[1] , Mason Andrew Kass[3]

Corresponding author: boxzuo@cug.edu.cn

Boxin Zuo*[1], Xiangyun Hu[2], Lizhe Wang[1] , Yi Cai[1] , Mason Andrew Kass[3]

Corresponding author: boxzuo@cug.edu.cn

**The PDF file includes:**

**Text S1 to S3**

**Figures. S1 to S10**

**Table S1 & S2**

**Other Supplementary Materials for this manuscript include the following:**

**Movies S1 to S11**

**Text S1:** ***Solving equivalent electrical current with orbital magnetic observation data.***

With the orbital observation data $B$ to invert the remanent magnetization $\mathbf{M}_{rem}$ is a typical geophysical invert problem. The potential field $\phi$ of the whole 3-D space can be solved in the process.

$$-\nabla\cdot\left(\mu_0(\nabla\phi-\mathbf{M}_{rem})\right)=0 \qquad (\text{S1a})$$
$$B=\mu_0\nabla\phi \qquad (\text{S1b})$$

Then, 3-D magnetic field can be calculated with the potential field $\phi$ and $\mathbf{M}_{rem}$.

$$\mathbf{B}=\mu_0(\nabla\phi+\mathbf{M}_{rem}) \qquad (\text{S2})$$

The equivalent current $\mathbf{J}_{equ}$ is estimated with the remanent magnetization $\mathbf{M}_{rem}$.

$$\mathbf{J}_{equ}=\nabla\times\mathbf{M}_{rem} \qquad (\text{S3})$$

Composing above equations, the actual toroidal current exhibits linear relationship with the equivalent current $\mathbf{J}_{equ}$:

$$\mathbf{J}_{act}\cdot\chi_{trm}=\mathbf{J}_{equ} \qquad (\text{S4})$$

The solved current model $\mathbf{J}_{equ}$ is also a 3-D vector field, which direction distribution is coincident with the actual current $\mathbf{J}_{act}$ with a linear coefficient $\chi_{trm}$.

**Text S2:** ***Magnetizing process with actual electrical currents*** $\mathbf{J}_{act}$

$\mathbf{J}_{act}$ is the actual toroidal current flows in the lunar crust in the ancient time. It produces the poloidal magnetic field $\mathbf{B}_{mgz}$ (in units of T )that magnetizing the lunar crust with susceptibility $\chi_{cst}$. Is the permeability of free space and $\mu_{cst}=\mu_0\left(1+\chi_{cst}\right)$ (in units of H/m) .

$$\nabla\times\mathbf{B}_{mgz}=\mu_{cst}\mathbf{J}_{act} \qquad (\text{S5})$$

We estimate the magnetizing field strength $\mathbf{H}_{mgz}$ with:

$$\mathbf{H}_{mgz} = \mathbf{B}_{mgz} / \mu_{cst} \tag{S6}$$

Then, the thermoremanent magnetization $\mathbf{M}_{mgz}$ (in units of A/m) is estimated with the $\chi_{trm}$ is the thermoremanence susceptibility (Wieczorek et al, 2012).

$$\mathbf{M}_{mgz} = \chi_{trm} \mathbf{H}_{mgz} \tag{S7}$$

In the synthetic experiment, the $\chi_{trm}$ is set to the Mare basalts rock type 1.58E-3.

**Text S3:** ***Electrical current produced by the magma fluids flowing in the ancient core field***

$$\mathbf{J} = \sigma(\mathbf{E} + \mathbf{v} \times \mathbf{B}) \tag{S8}$$

Ohm's Law for Moving Conductors in a Magnetic Field is utilized. $\mathbf{J}$ is the current density, $\sigma$ is the electrical conductivity, $\mathbf{E}$ is the electric field, $\mathbf{v}$ is the velocity of the magma fluid, and $\mathbf{B}$ is the magnetic field. This equation describes the current density in the moving magma fluids within the ancient core field.

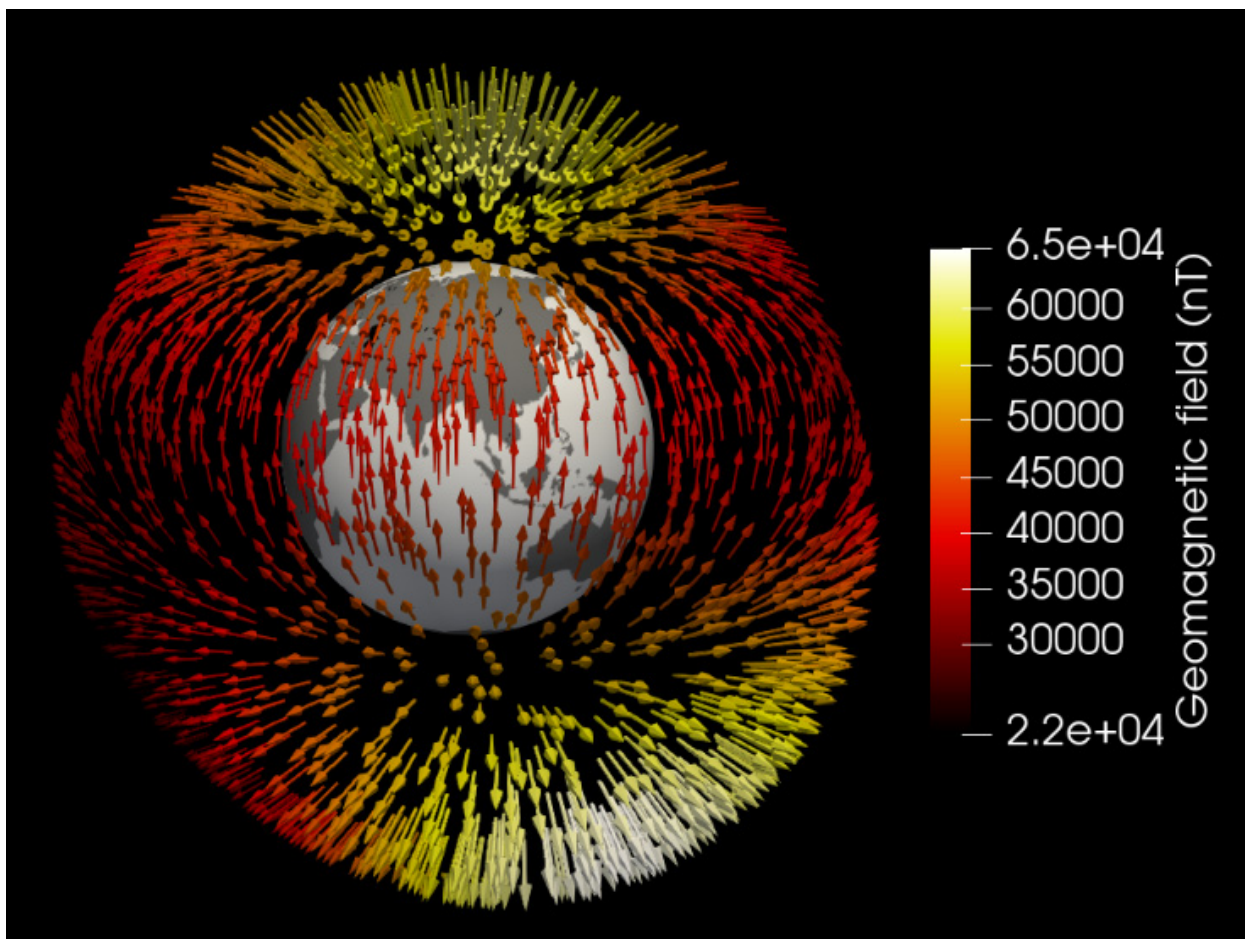

**Figure S1. 3D View of the 30km Altitude Geomagnetic Field Vector Data.**

This visualization is based on the IGRF-13 model, showcasing a global dipole field feature. A scaled-down model of Earth is included within the visualization to indicate directional orientation.

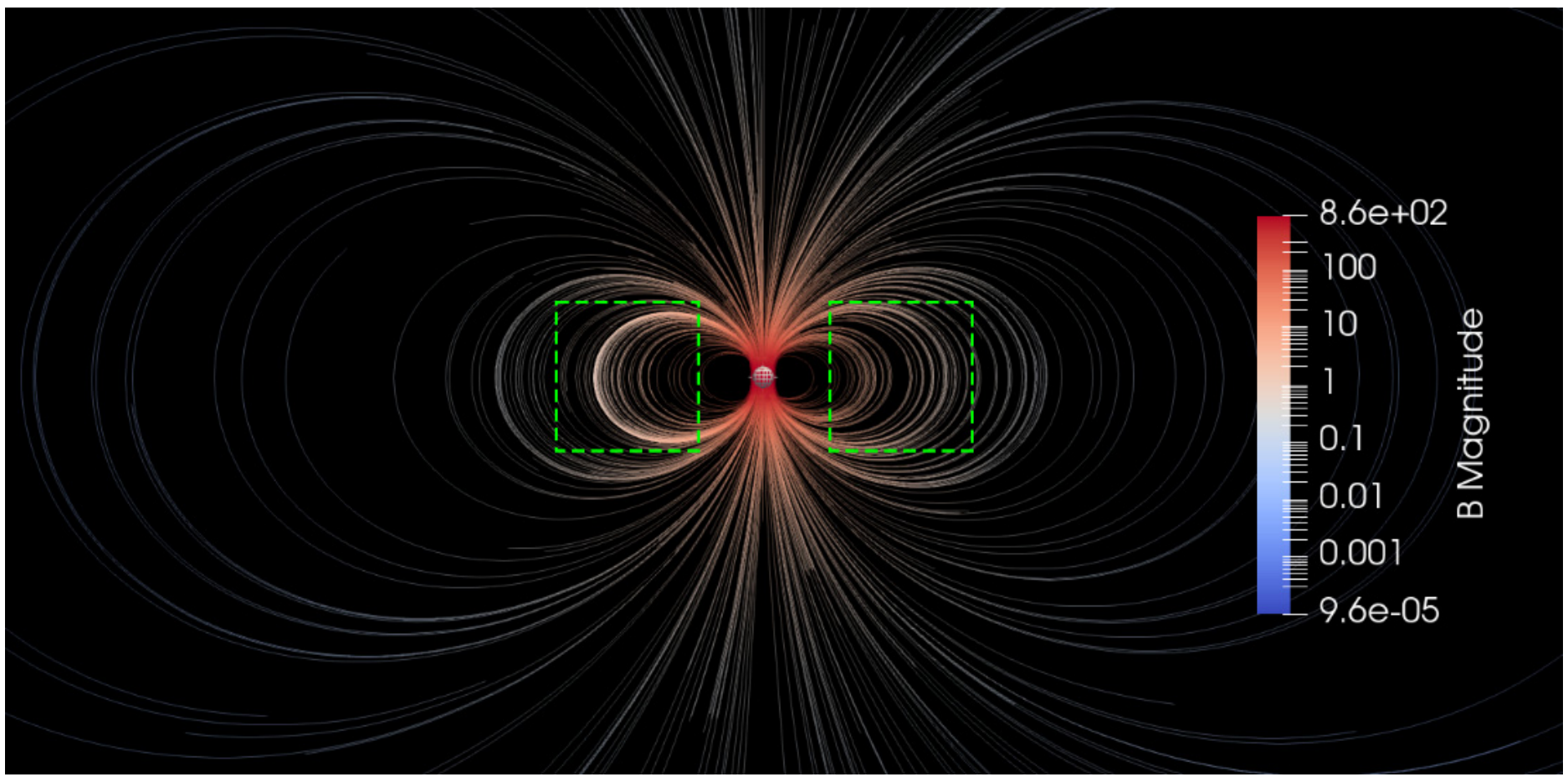

**Figure S2. Vertical Section of a Magnetic Dipole Field.** Green rectangles highlight areas where a reversal in the direction of the magnetic field occurs near the dipole source. Although the magnetic field is symmetric, the uneven distribution of sampling points results in an irregular pattern of magnetic field lines in the diagram.

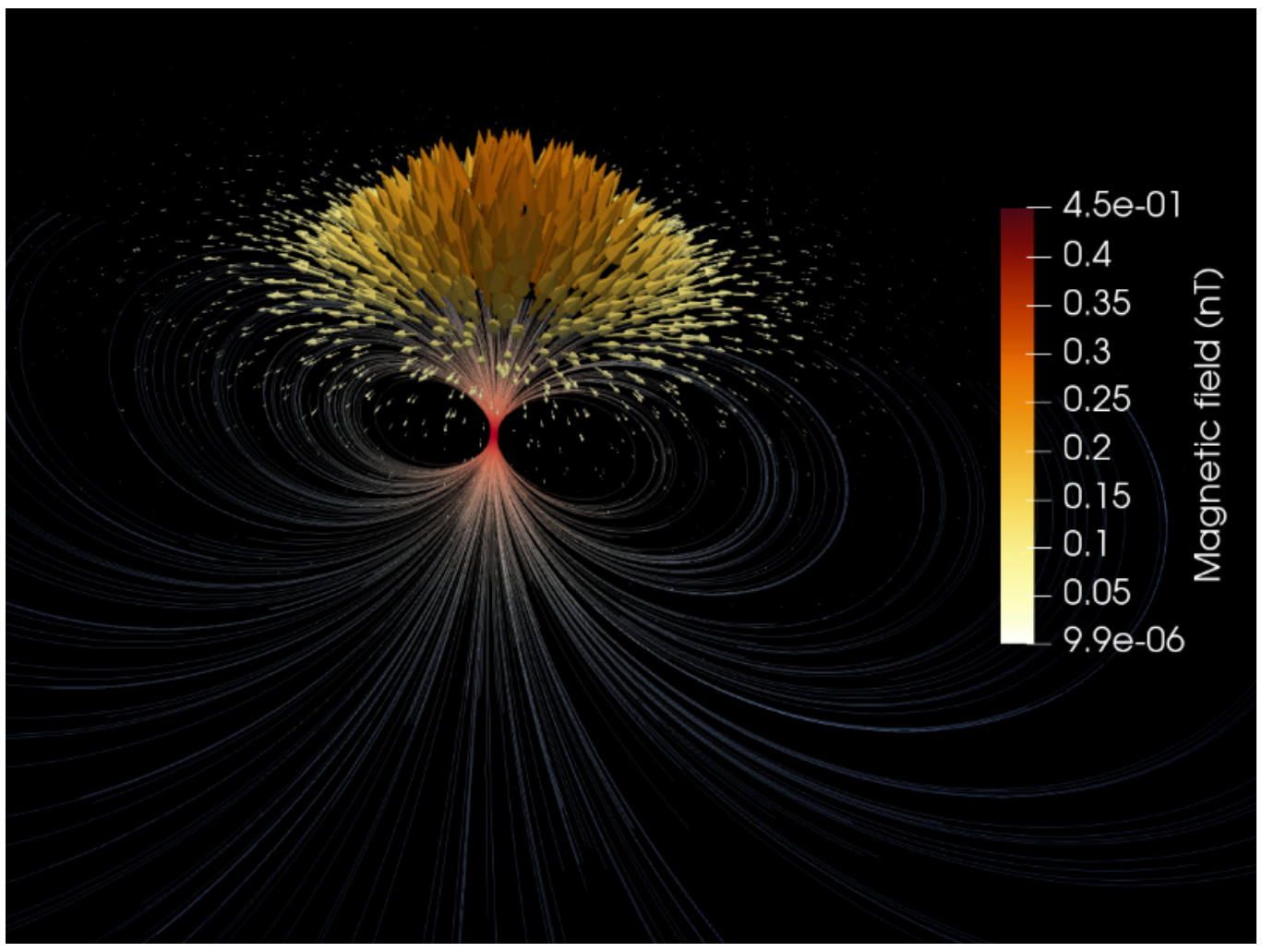

**Figure S3. Magnetic Field Distribution on a Horizontal Plane Above the Dipole.** This diagram illustrates the spatial arrangement of the magnetic field lines as observed in a horizontal section positioned directly above the dipole source.

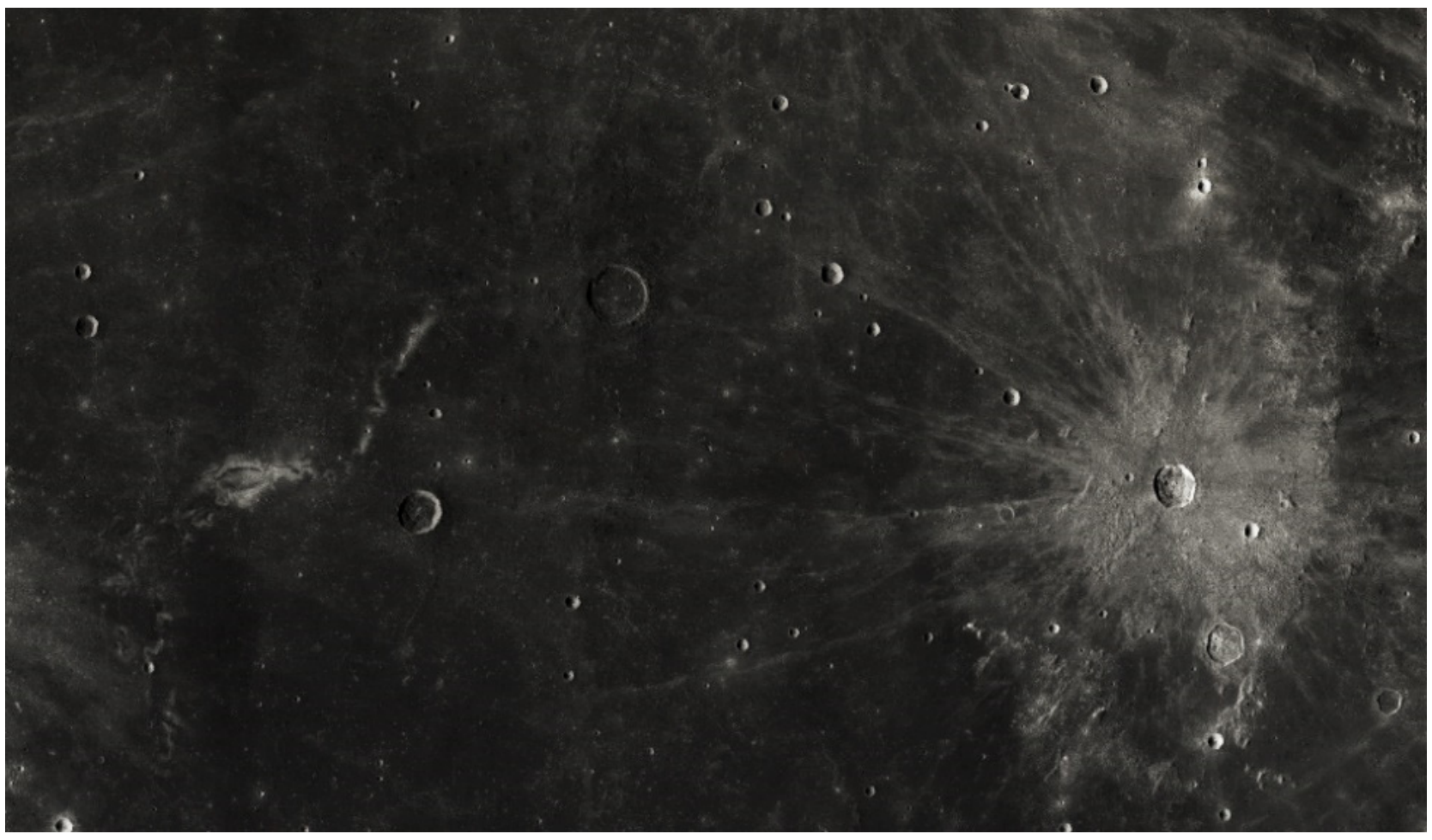

**Figure S4. High Albedo Optical Lunar Swirl and Ejected Plagioclase-Rich Branch in Oceanus Procellarum.**

This figure showcases a lunar swirl with high albedo characteristics, accompanied by a plagioclase-rich branch of ejected material, located in the Oceanus Procellarum region.

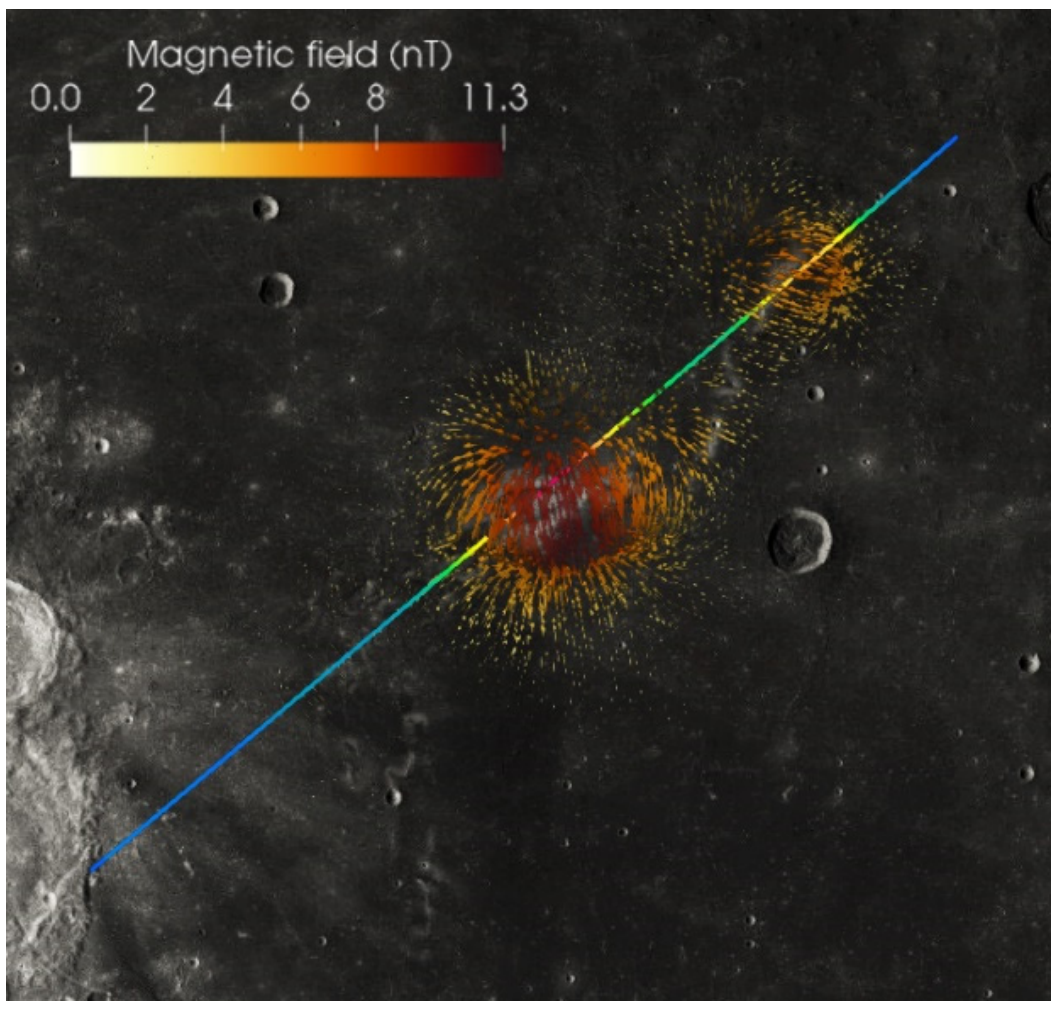

**Figure S5. Observed Magnetic Anomaly Vector Above Reiner-Gamma Region.**

This figure displays 3-D arrows representing the magnetic anomaly vectors above the Reiner-Gamma region. The colored line indicates the position of the data slice referenced in Fig. 5a, providing a spatial context for the observed anomalies.

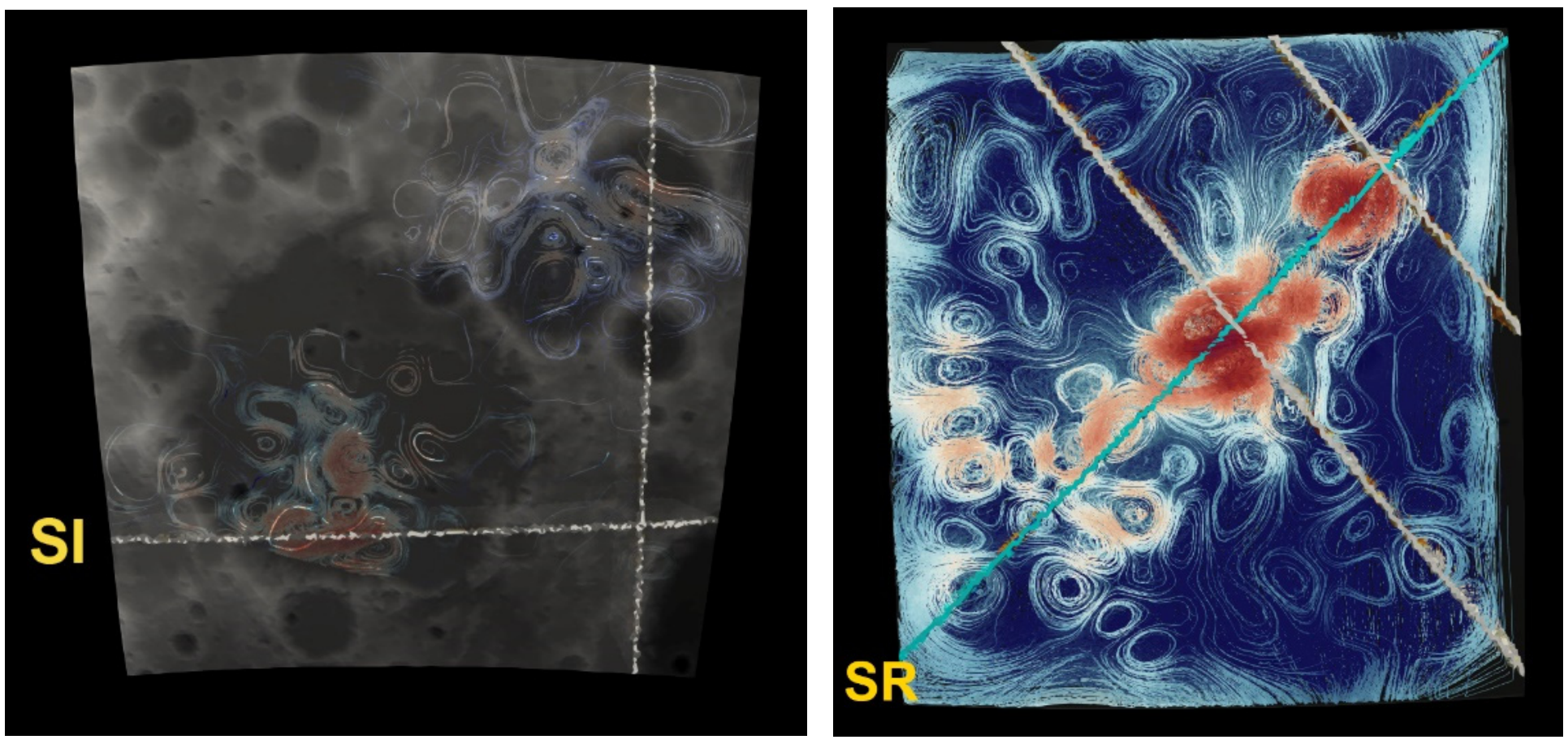

**Figure S6. Section Positions of the 3-D Density Inverse Model in Ingenii and Reiner-Gamma Regions.**

This diagram illustrates the locations of cross-sectional views within the 3-D density inverse models for the Ingenii and Reiner-Gamma regions.

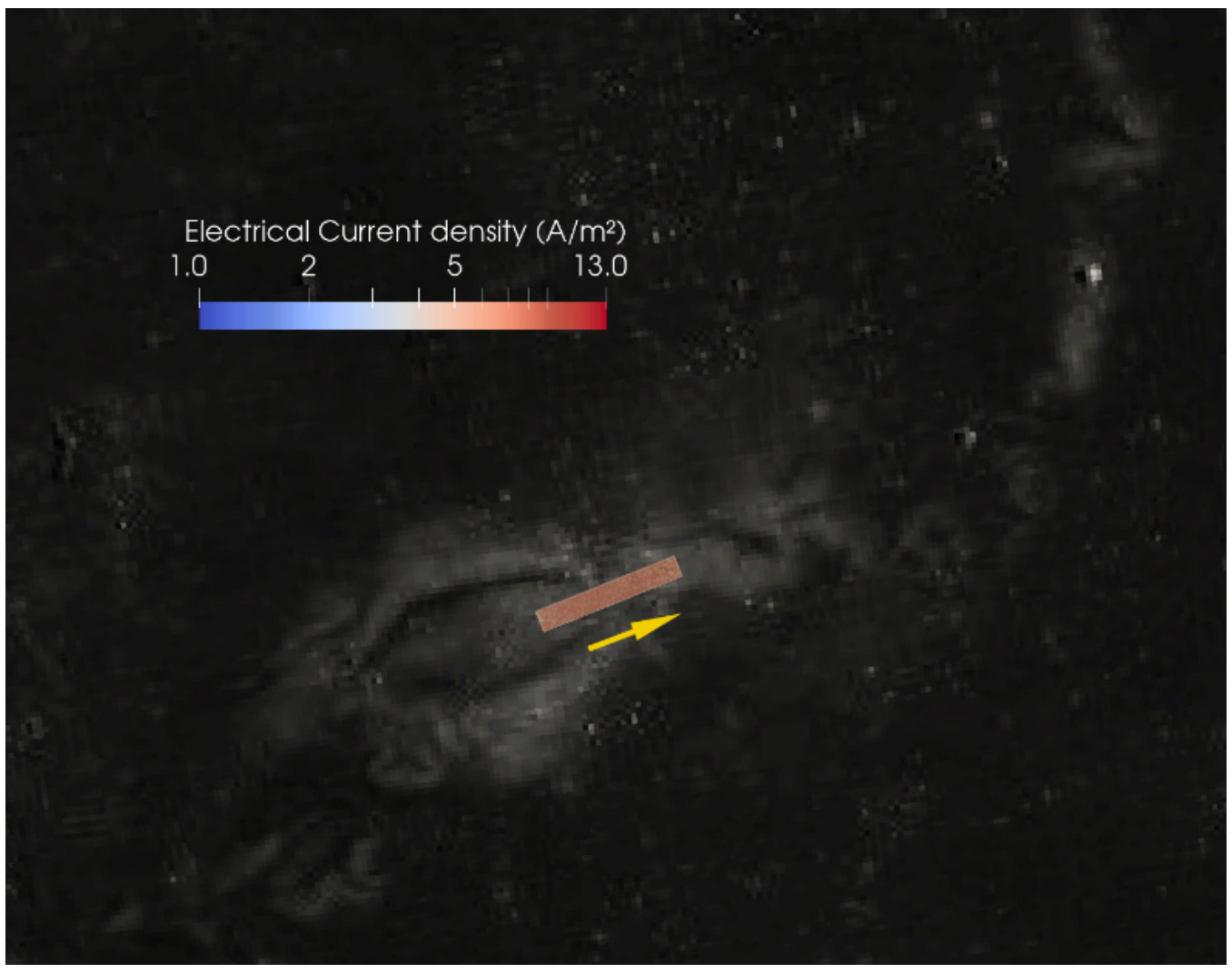

**Figure S7. Electrical Current Source Conductivity to the Crust.**

This diagram highlights the area of the electrical current source beneath the crust, delineated by the rectangle. The yellow arrow indicates the direction of the current flow.

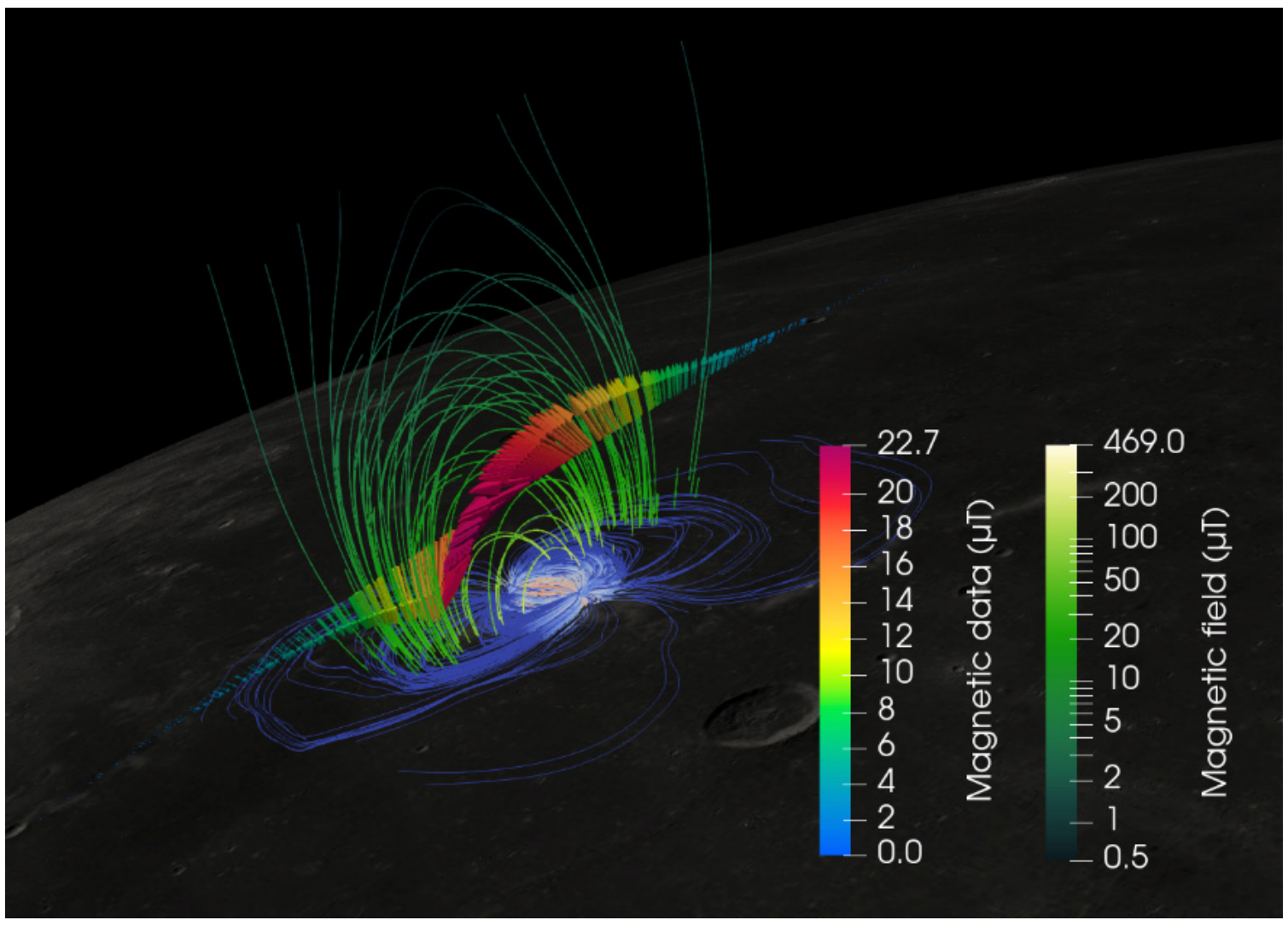

**Figure S8. Magnetizing Field of the Synthetic Example.**

The green line illustrates the magnetic field along the slice, while colored arrows represent observational data from an orbital position, providing a comprehensive view of the field's behavior in the synthetic model.

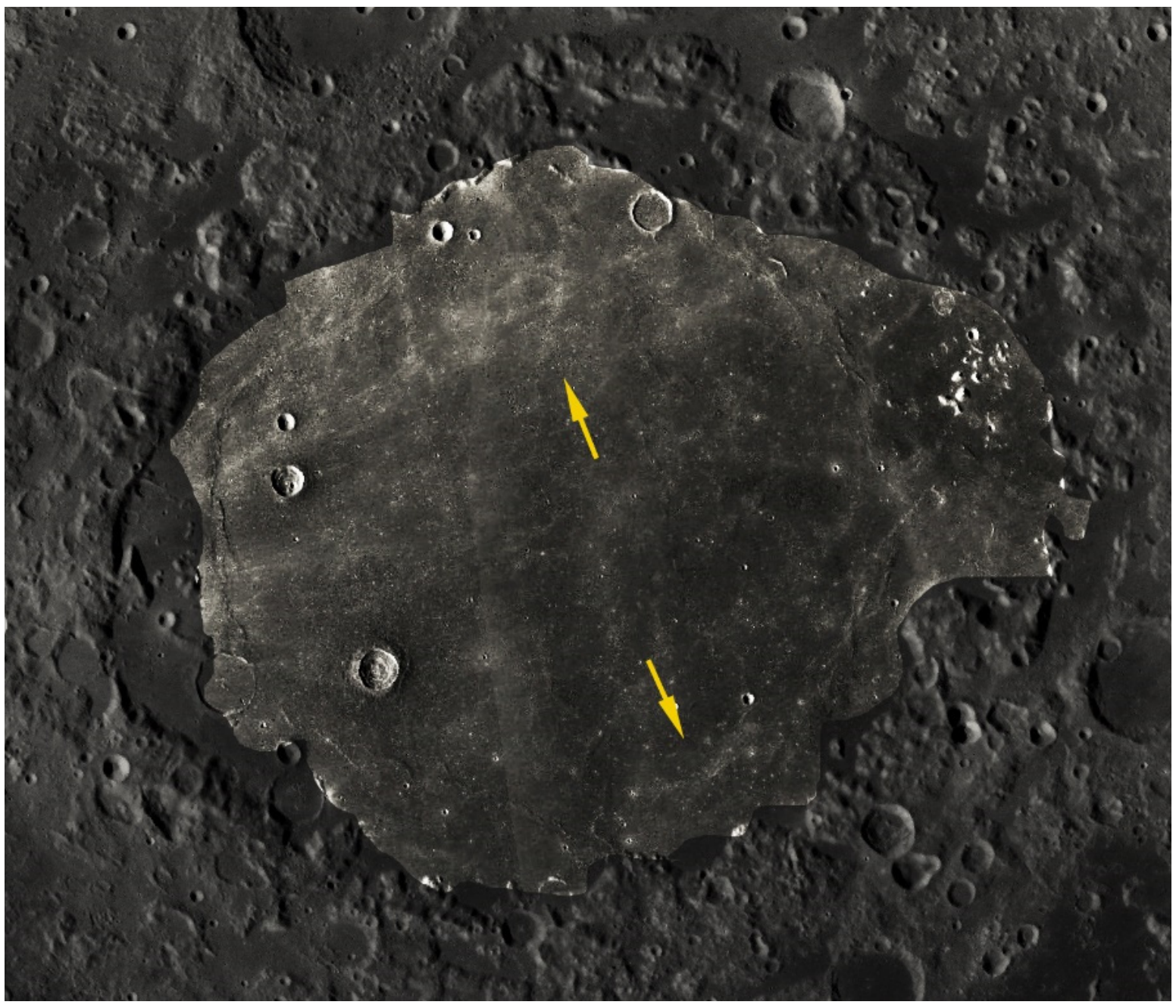

**Figure S9 Enhanced Optical Image of Crisium Basin.** This image focuses on the centrally enhanced part of the Crisium basin, highlighting optical anomaly features and details within the region.

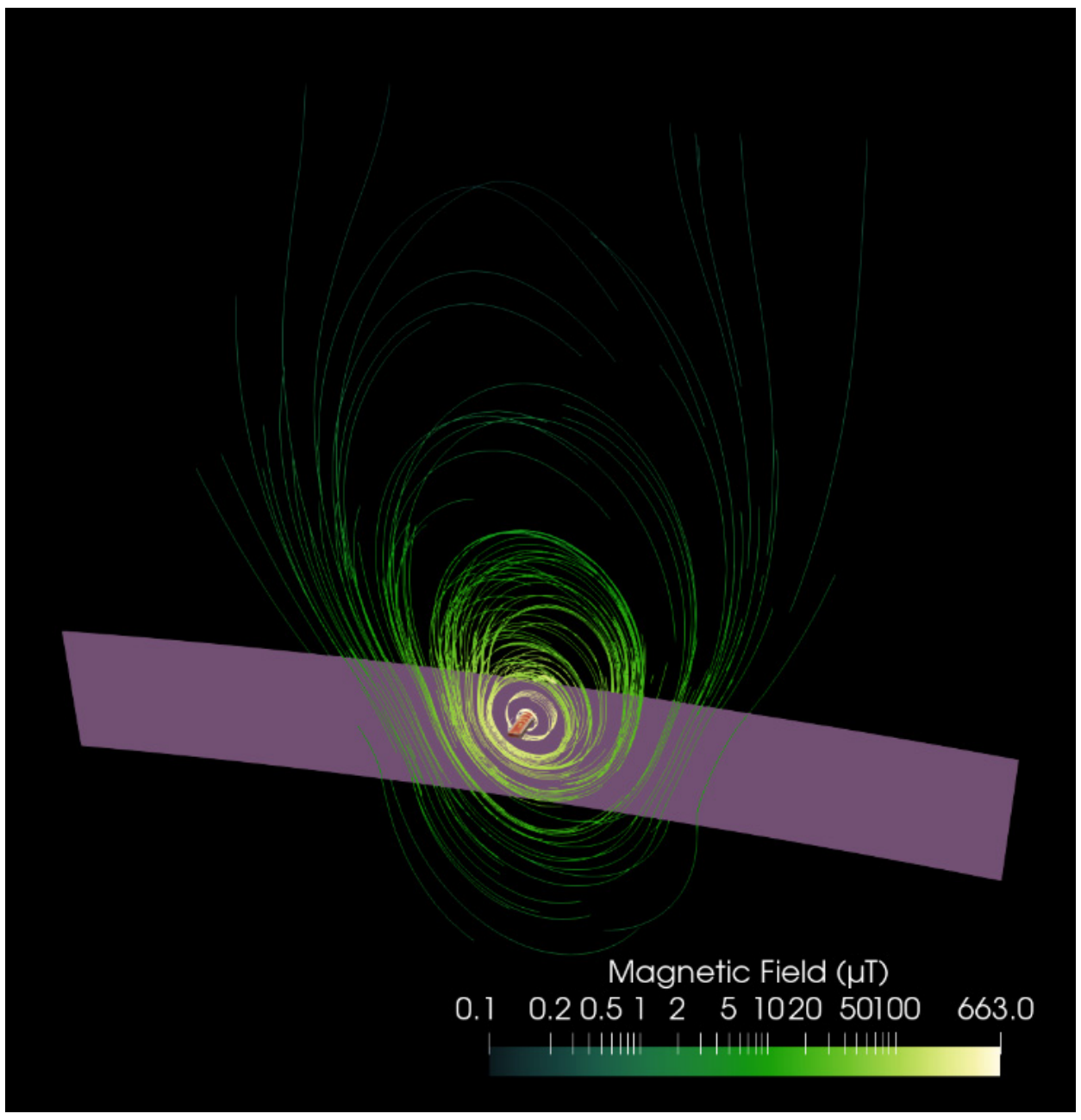

**Figure. S10. Slice of the Magnetic Field Induced by the Electrical Current Source.** This visualization represents a cross-sectional view of the magnetic field generated by an underlying electrical current source

**Table S1** Thermoremanent magnetization range of rocks in the studies region.

(Referring to Wieczorek 2012)

| Regions | Rock type | Density $kg/m^3$ | Possible $\chi_{TRM}$ Range (SI) |
|---|---|---|---|
| **Mendel-Ryberg (Basin) No Swirls** | ➢ Mafic impact-melt breccias<br>➢ Granulitic breccias<br>➢ Mare basalts | 2877 | $(13.62, 0.53) \times 10^{-3}$ |
| **Humboldtianum (Basin) No Swirls** | ➢ Mafic impact-melt breccias<br>➢ Granulitic breccias<br>➢ Mare basalts | 2858 | $(13.62, 0.53) \times 10^{-3}$ |
| **Crisium (Basin) Weak Swirls** | ➢ Mafic impact-melt breccias<br>➢ Granulitic breccias<br>➢ Mare basalts | 2945 | $(13.62, 0.53) \times 10^{-3}$ |
| **Leibnitz (Basin) With Swirls** | ➢ Mafic impact-melt breccias<br>➢ Granulitic breccias<br>➢ Mare basalts | 3185 | $(13.62, 0.53) \times 10^{-3}$ |
| **Ingenii (Mare) With Swirls** | ➢ Mare basalts | 3057 | $(5.72, 0.53) \times 10^{-3}$ |
| **Marginis (Mare) With Swirls** | ➢ Mare basalts<br>➢ Pristine feldpathic highland rocks | 3036 | $(5.72, 0.18) \times 10^{-3}$ |
| **Reiner- Gamma (Mare) With Swirls** | ➢ Mare basalts | 2780 | $(5.72, 0.53) \times 10^{-3}$ |

**Table S2** Convergence rate and errors of the inversions

| Regions | Magnetic field | Gravity field |
|---|---|---|
| **Mendel-Ryberg (Basin) No Swirls** | 2.9%<br>0.14 nT | 4.6%<br>22 mGal |
| **Humboldtianum (Basin) No Swirls** | 3.7%<br>1.2 nT | 2.1%<br>9.8 mGal |
| **Crisium (Basin) Weak Swirls** | 3.6%<br>0.38 nT | 11.1%<br>64 mGal |
| **Leibnitz (Basin) With Swirls** | 8%<br>0.4 nT | 1.9%<br>2.7 mGal |
| **Ingenii (Mare) With Swirls** | 6%<br>1.1 nT | 1.2%<br>2.5 mGal |
| **Marginis (Mare) With Swirls** | 7.2%<br>0.6 nT | 3.1%<br>4.3 mGal |
| **Reiner- Gamma (Mare) With Swirls** | 10.6%<br>1.2 nT | 1.5%<br>2.5 mGal |

**Movie**

All movies are cited in the captions of figures, so the captions of them are as same as the corresponding captions. The sizes of movies are too large to upload in the system, so they are uploaded online, as following:

**Movie S1.**
https://doi.org/10.6084/m9.figshare.26461630

**Movie S2.**
https://doi.org/10.6084/m9.figshare.26463373

**Movie S3.**
https://doi.org/10.6084/m9.figshare.26463673

**Movie S4.**
https://doi.org/10.6084/m9.figshare.26463805

**Movie S5.**
https://doi.org/10.6084/m9.figshare.26463901

**Movie S6.**
https://doi.org/10.6084/m9.figshare.26464048

**Movie S7.**
https://doi.org/10.6084/m9.figshare.26464183

**Movie S8.**
https://doi.org/10.6084/m9.figshare.26464681

**Movie S9.**
https://doi.org/10.6084/m9.figshare.26464828

**Movie S10.**
https://doi.org/10.6084/m9.figshare.26464930

**Movie S11.**
https://doi.org/10.6084/m9.figshare.26465017